\RequirePackage{fix-cm}
\documentclass[smallextended]{svjour3}       
\smartqed  
\usepackage{graphicx}
\usepackage{color}
\usepackage{hyperref}
\usepackage{latexsym}
\usepackage{amssymb}
\usepackage{amsfonts}
\usepackage{amsmath}
\usepackage{bm}
\begin{document}

\title{The key role of magnetic fields in binary neutron star mergers}

\titlerunning{Magnetic fields in BNS mergers}      

\author{Riccardo Ciolfi}

\institute{R. Ciolfi \at
              INAF--Osservatorio Astronomico di Padova, Vicolo dell'Osservatorio 5, I-35122 Padova, Italy \\
              INFN--Sezione di Padova, Via Francesco Marzolo 8, I-35131 Padova, Italy \\
              \email{riccardo.ciolfi@inaf.it}
}

\date{Received: date / Accepted: date}

\maketitle

\begin{abstract}

The first multimessenger observation of a binary neutron star (BNS) merger in August 2017 demonstrated the huge scientific potential of these extraordinary events. This breakthrough led to a number of discoveries and provided the best evidence that BNS mergers can launch short gamma-ray burst (SGRB) jets and are responsible for a copious production of heavy r-process elements. On the other hand, the details of the merger and post-merger dynamics remain only poorly constrained, leaving behind important open questions.
Numerical relativity simulations are a powerful tool to unveil the physical processes at work in a BNS merger and as such they offer the best chance to improve our ability to interpret the corresponding gravitational wave (GW) and electromagnetic emission. Here, we review the current theoretical investigation on BNS mergers based on general relativistic magnetohydrodynamics simulations, paying special attention to the magnetic field as a crucial ingredient.
First, we discuss the evolution, amplification, and emerging structure of magnetic fields in BNS mergers. Then, we consider their impact on various critical aspects: (i) jet formation and the connection with SGRBs, (ii) matter ejection, r-process nucleosynthesis, and radiocatively-powered kilonova transients, and (iii) post-merger GW emission. 
\keywords{
Binary Neutron Stars \and Gamma-ray Bursts \and Gravitational Wave Sources \and Magnetohydrodynamical Simulations 
}

\end{abstract}

\section{Introduction}
\label{intro}

The merger of two neutron stars (NSs) in a binary system is an event of the utmost importance for astrophysics and fundamental physics.
Binary neutron star (BNS) mergers are among the most promising sources of gravitational waves (GWs) and, at the same time, they are responsible for a variety of electromagnetic (EM) signals covering the whole spectrum. 
The extremely rich and complementary information from these different channels offers a unique opportunity to shed light on the unknown behaviour of matter at supranuclear densities, encoded in the NS equation of state (EOS), the formation channels of compact binary systems across cosmic time, the origin of heavy elements in the Universe, and much more (e.g., \cite{Faber2012,Baiotti2017,Paschalidis2017,Ciolfi2018,Duez2019,Ciolfi2020c} and refs.~therein). 

Besides their strong GW emission, BNS mergers have long been considered the leading scenario to explain the phenomenology of short gamma-ray bursts (SGRBs) \cite{Paczynski1986,Eichler1989,Narayan1992,Barthelmy2005a,Fox2005,Gehrels2005,Berger2014}. 
However, while the association of long-duration GRBs with core-collapse supernovae has been firmly established thanks a number of combined detections collected over the last two decades (the first case being SN\,1998bw \cite{Galama1998}), the BNS merger origin of SGRBs has remained only hypothetical until the recent GW detections started to offer a solid chance of proving it. 

Furthermore, BNS mergers represent an ideal site for r-process nucleosynthesis, which is responsible for the production of heavy elements \cite{Lattimer1974,Symbalisty1982,Eichler1989,Meyer1989}.  
The radioactive decay of these elements is expected to produce a potentially observable ``kilonova'' optical/infrared transient \cite{Li1998,Rosswog2005,Kulkarni2005,Metzger2010} that could accompany the GW signal from the merger and thus reveal the presence of newly-synthesized heavy elements. 
A candidate kilonova signature was already identified in the afterglow of GRB 130603B \cite{Tanvir2013,Berger2013}. Nevetheless, the evidence was not strong enough to establish a solid link among kilonovae, SGRBs, and compact binary mergers. 

The first multimessenger observation of a BNS merger in August 2017 allowed us to confirm all of the above expectations.\footnote{We note that while the BNS hypothesis remains favoured, the possibility of a NS-black hole system cannot be completely ruled out (e.g., \cite{LVC-170817properties,Hinderer2019}).}
This represented not only the first detection of a GW signal from such a system \cite{LVC-BNS}, but also the first combined detection of GW and EM signals from the same source \cite{LVC-MMA}. 
The analysis of the GW emission during the last orbits of inspiral up to merger allowed us to place new constraints on the NS EOS (e.g., \cite{LVC-170817properties,Raithel2019} and refs.~therein) and to estimate the luminosity distance of event. The latter estimate, combined with the redshift inferred from the host galaxy via EM observations, led to the first GW-based measurement of the Hubble constant \cite{LVC-Hubble}. 
The association with the SGRB named GRB 170817A \cite{LVC-GRB,Goldstein2017,Savchenko2017} showed that BNS mergers can produce SGRBs \cite{LVC-GRB,Goldstein2017,Savchenko2017,Troja2017,Margutti2017,Hallinan2017,Alexander2017,Mooley2018a,Lazzati2018,Lyman2018,Alexander2018,Mooley2018b,Ghirlanda2019}. Moreover, this was the first SGRB to be observed off-axis by $15^\circ\!-\!30^\circ$, which represented a unique opportunity to study the angular structure of the corresponding jet (e.g., \cite{Mooley2018b,Ghirlanda2019}). 
Finally, the optical/infrared signal AT\,2017gfo, that was crucial to pinpoint the host Galaxy, was clearly identified via photometric and spectroscopic analysis as a radioactively-powered kilonova, demonstrating the presence of r-process nucleosynthesis and the production of heavy elements in BNS mergers (e.g., \cite{Arcavi2017,Coulter2017,Valenti2017,Pian2017,Smartt2017,Kasen2017}; see also, e.g., \cite{Hotokezaka2018,Metzger2019LRR,Siegel2019} for recent reviews). 

This was certainly a breakthrough discovery, but a number of open questions remain, in part due to the poorly constrained post-merger dynamics.
The collected information suggests that the merger resulted in the formation of a metastable massive NS, which eventually collapsed into a black hole (BH). However, it was not possible to establish how long the metastable remnant was able to survive.
As a consequence, the SGRB could have been powered before the collapse, implying a massive NS central engine \cite{Zhang2001,Gao2006,Metzger2008}, or after the collapse, which would instead point to an accreting BH as the power source \cite{Eichler1989,Narayan1992,Mochkovitch1993} (see, e.g., \cite{Paschalidis2017,Ciolfi2018,Nathanail2018} for recent reviews).
A second open question of particular interest is about the origin and physical properties of the ejected material that was responsible for the kilonova. In order to match the observations, different combinations of ejecta components have been proposed, but the final answer is still matter of debate (e.g., \cite{Perego2017b,Metzger2018,Kawaguchi2018,Nedora2019}). 

Numerical relativity simulations represent the leading approach to unravel the physical mechanisms at play during a BNS merger and how the accompanying GW and EM emission depend on the properties of the source (including, e.g., the NS EOS). 
As such, these type of simulations can also shed light on SGRB central engines and the conditions for r-process nucleosynthesis and the production of a kilonova signal.
In this context, dramatic progress has been achieved since the first BNS merger simulations in full general relativity \cite{Shibata2000}, with the progressive inclusion of fundamental physical ingredients such as magnetic fields (e.g., \cite{Anderson2008,Liu2008,Giacomazzo2009,Giacomazzo2011,Rezzolla2011,Palenzuela2013,Kiuchi2014,Kiuchi2015,Endrizzi2016,Ruiz2016,Kawamura2016,Ciolfi2017,Ruiz2017,Ciolfi2019,Ruiz2019,Paschalidis2019,Tsokaros2019,Ciolfi2020a,Ruiz2020}) or composition dependent nuclear physics EOS and neutrino radiation (e.g., \cite{Sekiguchi2011,Neilsen2014,Sekiguchi2016,Kastaun2016,Radice2016,Foucart2016,Most2019}). 

In this review, we focus the attention on the pivotal role of magnetic fields in BNS mergers, as revealed by more than a decade of general relativistic magnetohydrodynamics (GRMHD) simulations. Moreover, we put most emphasis on the progress made over the last few years.\footnote{For a more general discussion on BNS mergers and the progress of numerical relativity simulations, we refer the reader to a number of other reviews on the subject, e.g. \cite{Faber2012,Baiotti2017,Paschalidis2017,Nathanail2018,Duez2019}.} 
First, we discuss the magnetic field of BNS systems and how it evolves during the merger process, its amplification and developing geometrical structure, and the impact on the properties of the (meta)stable massive NS remnant and surrounding environment (Section \ref{mag}). 
Next, we consider the connection with SGRBs, with special attention to the most recent results on both the accreting BH and the massive NS central engine scenarios, also in relation to the crucial case of GRB 170817A (Section \ref{sgrb}). 
In the following Section, we turn to discuss the different channels of matter ejection and in particular how the properties of the corresponding ejecta components and the associated kilonova emission are influenced by magnetic fields (Section \ref{ejectaKN}). In the same Section, we also consider different interpretations of the kilonova of August 2017.
Finally, we examine the potential impact of magnetic fields on the GW emission during the different merger phases (Section \ref{gw}) and devote the last Section to draw our conclusions (Section \ref{concl}).

\section{Magnetic field amplification and emerging structure}
\label{mag}

The present Section introduces the main features of the magnetic field structure and evolution in a BNS merger, as well as the impact of a strong magnetic field on the basic properties of the resulting massive NS remnant.

\subsection{The challenge of small-scale amplification}
\label{ampl}

Based on the observation of known BNS systems, the typical range of magnetic field strengths characterizing the two NSs before merger is $10^{10}\!-\!10^{12}$\,G (e.g., \cite{Lorimer2008}).\footnote{Isolated NSs with much stronger magnetic fields exist, up to the typical $10^{14}\!-\!10^{15}$\,G of magnetars. However, there is no evidence for such high magnetizations in BNSs.}
These values refer to the dipolar component of the field at the NS surface, while higher multiple components as well as the interior field could be stronger by one order of magnitude or more (e.g.,~\cite{Bilous2019,Rea2010}).
Such magnetization levels are considerable, but still insufficient to have any relevant effect on the merger and post-merger dynamics. However, the merger process itself is responsible for a strong magnetic field amplification up to levels that are dynamically important.
Magnetic energies of up to $10^{50}\!-\!10^{51}$\,erg and maximum field strengths $\gtrsim\!10^{16}$\,G can be achieved, making these systems the strongest magnets known in the Universe.

Magnetic field amplification occurs via a number of distinct mechanisms. As we discuss below, these include in particular the Kelvin-Helmholtz (KH) instability (e.g., \cite{Price2006,Anderson2008,Kiuchi2015}) and the magnetorotational instability (MRI; e.g., \cite{Balbus1991,Duez2006a,Siegel2013}), which dominate during and after merger, respectively, and act on small scales that can be extremely challenging to properly resolve with current computational resources (e.g., \cite{Kiuchi2018}). 

A certain degree of amplification may already occur prior to merger, even though a clear identification of the possible mechanism and its effects is still missing. 
This issue is addressed, e.g., in \cite{Ciolfi2017}, where the authors consider a variety of BNS merger models and observe an amplification by up to almost one order of magnitude in magnetic energy over the last one to few orbits.
Having ruled out a number of physical and numerical causes, they propose that the time-changing tidal deformations during inspiral, which induce fluid flows inside the two NSs, might be responsible for the observed amplification \cite{Ciolfi2017}. 
Nevetheless, further investigation is necessary for a solid assessment. 

\begin{figure*}
\begin{center}
  \includegraphics[width=0.95\textwidth]{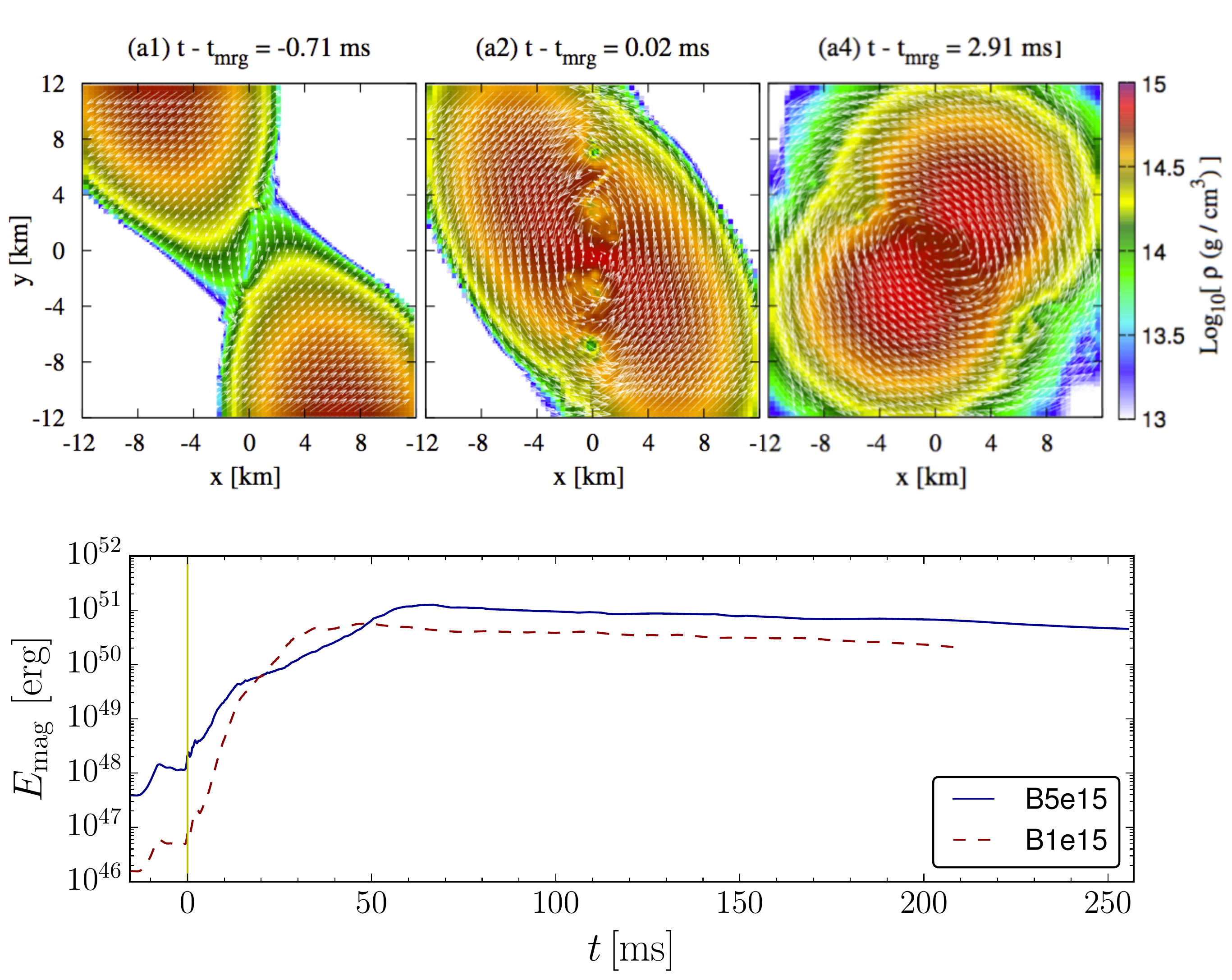}
\end{center}
\caption{{\it Top}\,(a): Equatorial snapshots of the KH instability developing when the two NSs come into contact (adapted from \cite{Kiuchi2015}). {\it Bottom}\,(b): Magnetic energy amplification and saturation for BNS merger simulations with two different initial magnetizations (from \cite{Ciolfi2020a}).}
\label{fig1}       
\end{figure*}

When the two NSs come into contact, the KH instability develops in the shear layer separating the two NS cores (Fig.~\ref{fig1}a). 
The toroidal component of the magnetic field, dragged along with the fluid motion, gets amplified by orders of magnitude reaching maximum field strengths up to $10^{15}\!-\!10^{17}$\,G (e.g., \cite{Price2006,Kiuchi2015}).
BNS merger simulations have shown that, irrespective of the initial field strength, magnetic energies of $\gtrsim\!10^{50}$\,erg can be reached in the few ms time window in which the KH instability is active \cite{Kiuchi2015,Kiuchi2018}.
At the same time, the most recent studies confirmed that even the highest resolution employed so far ($\sim\!12.5$\,m), with computational costs that are prohibitive for most numerical relativity groups, is still not enough to fully resolve the KH instability in BNS mergers \cite{Kiuchi2018}. This represents an important caveat for all current GRMHD simulations of such systems.

In order to account for the very high magnetizations expected in the post-merger phase despite the computational limitations, different approaches have been considered. 
The simplest one consists of imposing initial (pre-merger) magnetic field strengths of $\sim\!10^{14}\!-\!10^{16}$\,G, i.e. orders of magnitude higher than what suggested by the observations (e.g, \cite{Kiuchi2014,Ruiz2016,Ciolfi2017,Ciolfi2019}).
This allows to study the post-merger dynamics under the influence of strong magnetic fields, even though the quantitative results from the MHD evolution, at least in the first few tens of ms after merger, may not be fully reliable. 
The alternative is to adopt an effective treatment of the small-scale amplification that cannot be resolved. In this context, subgrid models have been proposed (e.g., \cite{Palenzuela2015,Giacomazzo2015}) where a modified induction equation accounts for the unresolved amplification via terms that depend on the fluid vorticity. Besides the assumption that vorticity directly relates to the small-scale dynamo, this method relies on closure scheme coefficients that need to be calibrated via very high resolution MHD simulations. 
A different possibility, currently under development and not yet applied to BNS mergers, is to perform explicit large-eddy MHD simulations, based on a more rigorous separation of the resolved and unresolved scales in combination with a subgrid model to be applied to the unresolved part \cite{Carrasco2020}. 
Finally, a further option that has been considered is to evolve the viscous hydrodynamics equations in substitution of the MHD equations \cite{Duez2004,Shibata2017a,Radice2017,Fujibayashi2020}, with the aim of capturing small-scale MHD effects while leaving aside any role of magnetic fields at larger scales. 
Also in this case, the approach depends on one or more parameters to be calibrated via very high resolution MHD simulations. 

Once a stable or metastable massive NS remnant has formed from the fusion of the two original NSs, KH instability is gradually substituted by other amplification mechanisms driven by the strong differential rotation of the new object, i.e. magnetic winding and the MRI. 
Magnetic winding arises from the fact that rotation is not uniform along poloidal magnetic field lines, resulting in amplification of the toroidal field which is linear with time.
The MRI is instead associated with modes that grow exponentially with time, with a much higher potential to amplify the field (e.g., \cite{Balbus1991,Duez2006a,Siegel2013}). The fastest growing mode of the MRI has a characteristic wavelength that scales with the magnetic field strength, i.e.~$\lambda_\mathrm{MRI}\!\sim\!2\pi B/\Omega\sqrt{4\pi \rho}$, where $B$, $\Omega$, and $\rho$ are the local field strength, angular velocity, and rest-mass density.
As a consequence, the strongest is the magnetic field, the better the MRI can be resolved with a given resolution. 

Recent GRMHD BNS merger simulations starting with maximum pre-merger magnetic field strengths as high as $10^{15}\!-\!10^{16}$\,G (as measured at the center of the two NSs by the Eulerian observer) and following the NS remnant evolution for $\approx\!100\!-\!250$\,ms \cite{Ciolfi2019,Ciolfi2020a} showed that, for the adopted resolution, already at $30\!-\!40$\,ms after merger the MRI is well resolved everywhere in the system except for the central region (cylindrical radius $\lesssim\!10$\,km) where the differential rotation profile has a positive angular velocity gradient and thus the MRI is anyways not expected to operate (see discussion in \cite{Ciolfi2019} and Section \ref{remnant}). 
Hence, while the previous MHD evolution is heavily underresolved (in particular the KH phase), the following one should be, at least qualitatively, reliable. 
Simulations like these cannot provide a solid quantitative link between the pre-merger system properties and the final outcome of the merger, but they can still be very useful to investigate the main features of strongly magnetized NS remnants \cite{Ciolfi2019,Ciolfi2020a}. 

One important aspect revealed by GRMHD simulations with a long enough post-merger evolution is the fact that magnetic field amplification is characterized by a physical saturation around magnetic energies of $10^{50}\!-\!10^{51}$\,erg (e.g.,~\cite{Kiuchi2018,Ciolfi2019,Ciolfi2020a}; see Fig.~\ref{fig1}b). 
This suggests that the maximum magnetic energy attainable in a BNS merger is limited to no more than a few times $10^{51}$\,erg. 

For metastable NS remnants, the MHD evolution and magnetic field amplification discussed above only hold as long as the remnant itself survives the collapse to a BH. 
When the collapse occurs, however, magnetic fields can be further amplified in the magnetized accretion disk around the BH, with the MRI maintaining the role of dominant amplification mechanism (e.g., \cite{Siegel2018}).

\subsection{Geometrical structure of the field}
\label{geom}

Current GRMHD simulations of BNS merger employ initial (pre-merger) magnetic field configurations that are superimposed by hand on a purely hydrodynamical solution for the BNS system under consideration (at the chosen initial separation, typically $\sim\!50$\,km).
This common approach is justified by the fact that even for the highest magnetizations considered (of order $10^{16}$\,G at the NS center for the Eulerian observer), magnetic fields represent only a small perturbation of the structure of the two NSs composing the binary.

Moreover, for simplicity, the initial magnetic field is in most (if not all) cases purely poloidal and dipolar. 
A more realistic configuration should also include an internal toroidal component of strength at least comparable to the poloidal one, as in the case of twisted-torus geometries (e.g., \cite{Ciolfi2013}), and might have a relevant contribution from higher multiples or significant deviations from axisymmetry. 
Nonetheless, the effects of employing a more general variety of initial magnetic field geometries remain to be investigated. 

An additional choice common to most magnetized BNS merger simulations is to impose that the magnetic field is initially confined to the NS interiors. 
This allows to avoid numerical issues that would arise in the very low density medium outside the two NSs, as the ideal MHD implementation commonly adopted is not able to handle high magnetic-to-fluid energy density ratios.
Different authors attempted to overcome this limitation by introducing a force-free treatment to be employed at very low densities (i.e., outside the two NSs) and continuously matched to the ordinary ideal MHD treatment of denser regions, either via an effective scheme mimicking the force-free condition (e.g., \cite{Ruiz2016}) or via a resistive GRMHD implementation with a chosen prescription for the electrical conductivity (e.g., \cite{Palenzuela2013,Ponce2014}). 
Based on the above approches, a number of simulations have been performed with initial dipolar fields extending outside the two NSs, allowing to study the interaction of the NS magnetospheres prior to merger and possible precursor EM signals (e.g., \cite{Palenzuela2013,Ponce2014}) or the effects of extended per-merger dipolar fields on the formation of incipient jets (e.g., \cite{Ruiz2016}).

\begin{figure*}
\begin{center}
  \includegraphics[width=0.92\textwidth]{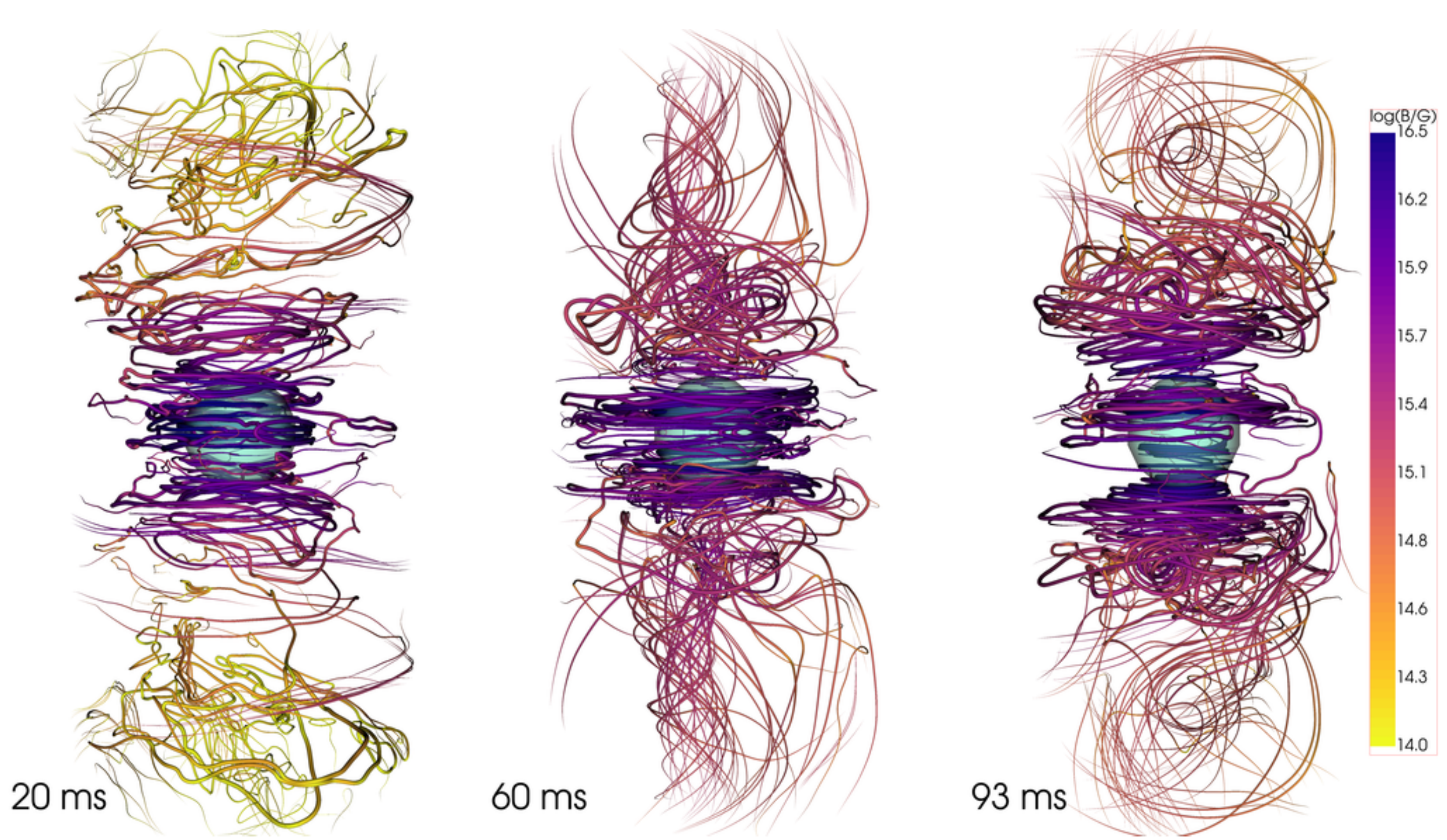}
\end{center}
\caption{Magnetic field structure at 20, 60, and 93\,ms after a BNS merger (from \cite{Ciolfi2019}). As a scale reference, a cyan sphere of 10\,km radius is placed at the remnant center of mass.}
\label{fig2}       
\end{figure*}

The merger process has a major impact on the magnetic field configuration that characterizes the massive (stable or metastable) NS remnant. 
A strong toroidal component is always produced, in particular on the equatorial region, due to the KH instability, magnetic winding, and the MRI.
Within a few tens of ms after merger, MHD turbulence redistributes the magnetic energy, leading to comparable toroidal and poloidal field strengths. 
In the inner equatorial region, toroidal fields remain dominant (Fig.~\ref{fig2}). 

Prior to the eventual collapse of the massive NS remnant (if any), the strong differential rotation within the bulk of the latter would tend to twist poloidal magnetic field lines along the spin axis and build up a nearly axisymmetric helical geometry, which would then represent a favourable condition to accelerate a collimated outflow and possibly an incipient SGRB jet (e.g., \cite{Shibata2011,Kiuchi2012,Siegel2014,Ciolfi2020a}; see Section \ref{magnetar}).
However, the emergence of such a structure requires a certain amount of time, which depends on the interplay with the opposing baryon-loaded environment surrounding the bulk of the NS remnant \cite{Ciolfi2020a}. 
If no extended helical structure has formed along the remnant spin axis by the time differential rotation in the core has been removed (typically a few hundreds of ms), it will either take a much longer time or not be able to form at all \cite{Ciolfi2020a}. 
Therefore, there is no guarantee that a well defined helical geometry is present when a metastable NS remnant collapses to a BH, even for a rather long-lived remnant \cite{Ciolfi2017,Ciolfi2019,Ciolfi2020a}. 
On the other hand, stable NS remnants have time to evolve on the long term ($\gg$1 s) towards a uniformly rotating and strongly magnetized object that may end up having a significant (if not dominant) global dipolar magnetic field of magnetar-like field strength \cite{Ciolfi2019,Ciolfi2020a}. 

After a metastable NS remnant has collapsed to a BH, the material that is still bound to the system and not immediately swallowed by the BH forms a thick torus-shaped accretion disk. 
Along the spin axis, where the centrifugal support is lacking, most of the material in the vicinity of the BH (up to $\sim\!200$\,km \cite{Ciolfi2017}) is rapidly accreted, creating a funnel of much lower density and half-opening angle of $\sim\!30^\circ$ (e.g., \cite{Rezzolla2011,Kawamura2016}). 

In terms of magnetic field configuration, the disk is characterized by MHD turbulence and by a mixed toroidal-poloidal field, where the toroidal component remains generally dominant (e.g., \cite{Siegel2018}).
At the inner disk-funnel interface, a twister-like structure is produced \cite{Kawamura2016}.
Finally, inside the funnel, the inflow of material initially stretches the magnetic field lines along the radial direction \cite{Rezzolla2011}. 
Then, the same field lines are gradually wound-up under the influence of the differentially rotating accretion disk, building up a helical structure along the spin axis characterized by an increasing magnetic pressure gradient opposing to the gravitational pull. 
Depending on the initial poloidal magnetic field strength in the funnel, the inflow along the axis may eventually be halted and converted into outflowing motion (the very beginning of an incipient jet \cite{Ruiz2016}; see Section \ref{BH}).

\subsection{Magnetized neutron star remnants}
\label{remnant}

\begin{figure*}
\begin{center}
  \includegraphics[width=0.99\textwidth]{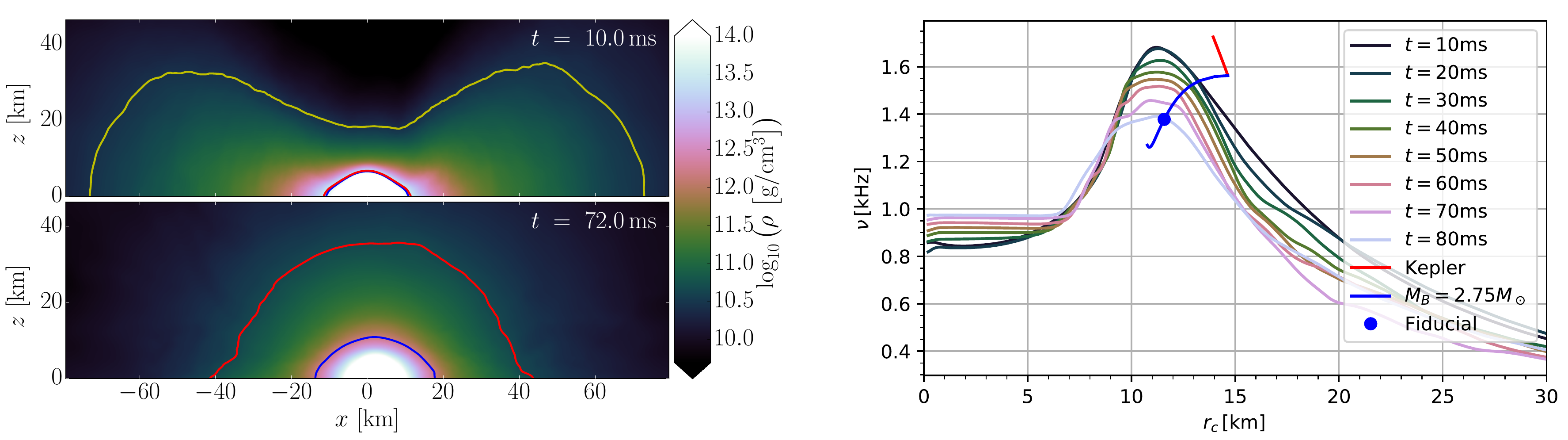}
\end{center}
\caption{{\it Left}\,(a): Meridional view of a magnetized massive NS remnant at 10 and 72 ms after merger (from \cite{Ciolfi2019}). In addition to the color-coded rest-mass density, three isodensity contours are shown that contain 92\% (blue), 93\% (red), and 98\% (yellow, only visible in the top panel) of the total baryon mass.
{\it Right}\,(b): Rotation profile evolution in the equatorial plane for the same model (from \cite{Ciolfi2019}). }
\label{fig3}       
\end{figure*}

Shortly after merger, the structure of a massive NS remnant is given by a spheroidal central core continuously attached to a torus-shaped outer envelope (see Fig.~\ref{fig3}a, top panel). In the literature, such a structure is often referred to as a massive NS surrounded by an ``accretion disk''. It should be noted, however, that in this case there is no distinction between a central object and a surrounding disk, nor a clearly defined accretion rate. 
The system is composed by a single deformed object in which the centrifugal support is gradually removed (see below) and, as a consequence, part of the mass tends to migrate inwards.  

As discussed in a number of studies (e.g., \cite{Shibata2005,Kastaun2015,Endrizzi2016,Kastaun2016,Hanauske2017,East2019} and refs. therein), the early rotation profile shows a slower spinning central region and a faster spinning outer layer, reaching the maximum angular velocity at cylindrical radii around $10\!-\!15$\,km
(see Fig.~\ref{fig3}b). At larger distances, the profile declines reproducing a nearly Keplerian behaviour, indicating that the outer part of the remnant is centrifugally supported. 
The time evolution exhibits a rapid consumption of the differential rotation in the bulk of the NS remnant due to the effective shear viscosity, with the maximum angular velocity descreasing and the central angular velocity slightly increasing (Fig.~\ref{fig3}b). Within a few hundreds of ms, uniform rotation is established in the remnant core, while a nearly Keplerian profile persists at larger distances (e.g., \cite{Ciolfi2020a}).
To compensate for the associated inward transport of angular momentum, the outer part of the remnant is expected to expand towards larger cylindrical radii (e.g., \cite{Fujibayashi2018}).

In parallel to the effects of angular momentum redistribution, the development of turbulence results in an effective pressure that pushes outwards material from the outer layers of the remnant, polluting the surrounding environment. 

While the above qualitative picture holds even for nonmagnetized BNS merger simulations, the inclusion of magnetic fields have a substantial impact on the NS remnant evolution.
As discussed in Section \ref{ampl}, differential rotation and turbulent fluid motions amplify the magnetic field until it becomes dynamically important. The corresponding build-up of magnetic pressure strongly enhances the outward push on the outer layers of the NS remnant, resulting in a mass outflow rate a factor of few to several higher (e.g., \cite{Ciolfi2017,Ciolfi2019}).
In this phase, the NS remnant is effectively expanding (see, e.g., the red isodensity contour in Fig.~\ref{fig3}a), turning most of the differential rotation energy being consumed into a positive binding energy variation and heat, while only a relatively small fraction ($\lesssim\!1\%$) turns into kinetic energy of the expanding material (see discussion in \cite{Ciolfi2019}). 

Since the magnetic field structure shortly after merger is rather disordered, the induced mass outflow from the outer layers of the NS remnant is, at least at early times, nearly isotropic \cite{Siegel2014}. As a consequence, the remnant gradually recovers a more isotropic density distribution (Fig.~\ref{fig3}a, bottom panel). 
Within a few to several tens of ms, the system consists of a central NS core surrounded by a cloud of material extending up to hundreds of km in all directions and characterized by a slow and continuous outflow (e.g., \cite{Ciolfi2019}).   
Pushed to larger and larger distances, part of this material can eventually become unbound and turn into a slow and potentially massive ejecta component (see Section \ref{ejectaKN}). 

As the evolution continues and as long as no collapse to a BH occurs, the above nearly isotropic condition could be altered by the gradual development of an ordered helical magnetic field structure along the spin axis, which would start accelerating the material in such direction. 
However, as discussed in Sections \ref{geom} and \ref{magnetar}, this behaviour may only apply to a limited fraction of BNS mergers \cite{Ciolfi2020a}.

\section{Jet formation and short gamma-ray bursts}
\label{sgrb}

The first multimessenger observation of a BNS merger in August 2017, coincident with the SGRB named GRB\,170817A, provided the long awaited smoking gun evidence that these systems can produce SGRBs \cite{LVC-GRB,Goldstein2017,Savchenko2017,Troja2017,Margutti2017,Hallinan2017,Alexander2017,Mooley2018a,Lazzati2018,Lyman2018,Alexander2018,Mooley2018b,Ghirlanda2019}. 
The analysis of the prompt gamma-ray signal of GRB\,170817A and the following multiwavelength afterglow radiation monitored for several months in the X-ray, optical/IR, and radio bands allowed to establish that a relativistic jet was launched by the merger remnant (\cite{Mooley2018b,Ghirlanda2019} and refs.~therein), in full agreement with the consolidated GRB paradigm (e.g., \cite{piran2004,Kumar2015}). Moreover, this GRB was observed off-axis by $15^\circ\!-\!30^\circ$ and the observed prompt gamma-ray emission, orders of magnitude less energetic than any other known SGRB, was not produced by the narrow and highly relativistic jet core pointing away from us (half-opening angle $\lesssim\!5^\circ$ and Lorentz factor $\Gamma\!\gtrsim10\!-\!100$), but rather by a mildly relativistic outflow along the line of sight (e.g., \cite{LVC-GRB,Lazzati2018,Mooley2018b,Ghirlanda2019}).\footnote{Such an outflow is naturally explained in terms of a ``structured jet'' composed by a highly collimated and energetic core surrounded by a less energetic wide-angle cocoon, where the latter is formed via the early interaction of the incipient jet with the baryon-polluted environment around the merger site (e.g., \cite{Lazzati2018}).}
The on-axis observer, on the other hand, would have found the properties of GRB\,170817A consistent with the known population of SGRBs (e.g., \cite{Salafia2019}).

This extraordinary SGRB allowed us to gain crucial insights on the ultimate angular structure of the escaping jet and its propagation across the interstellar medium. Nevertheless, the obtained constraints on the jet launching mechanism and the properties of central engine are very limited. 
In particular, the jet was launched in a time window between the merger and $\approx\!1.74$\,s later, which corresponds to the onset of the observed gamma-ray emission \cite{LVC-GRB}, but there is no compelling evidence on whether the BNS merger remnant, most like a metastable massive NS, collapsed to a BH within such time window or later. As a consequence, both an accreting BH \cite{Eichler1989,Narayan1992,Mochkovitch1993} and a massive NS \cite{Zhang2001,Gao2006,Metzger2008} represent viable central engines for GRB\,170817A.

The nature of the central engine of SGRBs and the associated jet launching mechanism represent a long-standing open problem (see, e.g., \cite{Ciolfi2018} for a recent review). 
BNS merger simulations in full general relativity offer a unique chance to study in detail the process of jet formation and to explore the different physical scenarios and, even though the final answers are still ahead of us, great progress has been made in this direction over the last decade. 
So far, most of this effort has been devoted to the case of a BH central engine \cite{Rezzolla2011,Kiuchi2014,Ruiz2016,Kawamura2016,Ruiz2020}, but the massive NS alternative has also recently started to receive a growing attention \cite{Ciolfi2017,Ciolfi2019,Ciolfi2020a}. 
 In the following, we discuss the two cases separately.

\subsection{Accreting black hole central engine}
\label{BH}

A spinning BH surrounded by a thick and massive ($\sim\!0.1\,M_\odot$) accretion disk is a likely outcome of a BNS merger and a promising candidate as a SGRB central engine.  
The highly super-Eddington accretion rates up to $\sim\!0.1\!-\!1\,M_\odot/$s and the typical accretion timescales of $\sim\!0.1\!-\!1$\,s may naturally explain the range of energies, luminosities, and durations of SGRBs (e.g., \cite{Eichler1989,Narayan1992,Mochkovitch1993}).
Moreover, such a system is characterized by an almost baryon free environment along the spin axis, at least within distances of a few hundred km, which represents a very favourable condition for producing a jet able to reach the required terminal Lorentz factor of $\Gamma\!\gtrsim\!10\!-\!100$.\footnote{Although we restrict the present discussion to BNS mergers, an accreting BH with the right properties to power a SGRB could also result from a NS--BH merger (see, e.g., \cite{Paschalidis2015} and refs.~therein).}

In terms of jet production, two main power sources have been proposed: neutrino--antineutrino annihilation near the poles of the BH (e.g., \cite{Mochkovitch1993,Ruffert1999,Aloy2005}) and energy extraction via strong magnetic fields (e.g., \cite{Blandford1977}).
While the results of recent simulations suggest that a neutrino-powered jet would not be powerful enough to explain a SGRB (e.g., \cite{Just2016,Perego2017a}), the idea that magnetic fields are the dominant driver remains viable and is currently favoured.
The most discussed mechanism within the latter paradigm is the Blandford--Znajek (BZ) one \cite{Blandford1977}, in which a spinning BH threaded by a strong magnetic field connected to the accretion disk can generate a Poynting-flux dominated outflow with power as high as \cite{Thorne1986}
\begin{equation}
L_\mathrm{BZ}\sim 10^{52} \Big{(}\frac{\chi}{0.5}\Big{)}^2  \Big{(}\frac{M_\mathrm{BH}}{2.5\,M_\odot}\Big{)}^2 \Big{(}\frac{B_\mathrm{BH}}{10^{16}\,\mathrm{G}}\Big{)}^2~\mathrm{erg/s}\, ,
\nonumber
\end{equation}
where $\chi$ and $M_\mathrm{BH}$ are the BH dimensionless spin and the BH mass, while $B_\mathrm{BH}$ is the characteristic magnetic field strength close to the BH. In this case, energy is directly extracted from the BH rotation via magnetic torque. 

\begin{figure*}
\begin{center}
  \includegraphics[width=0.94\textwidth]{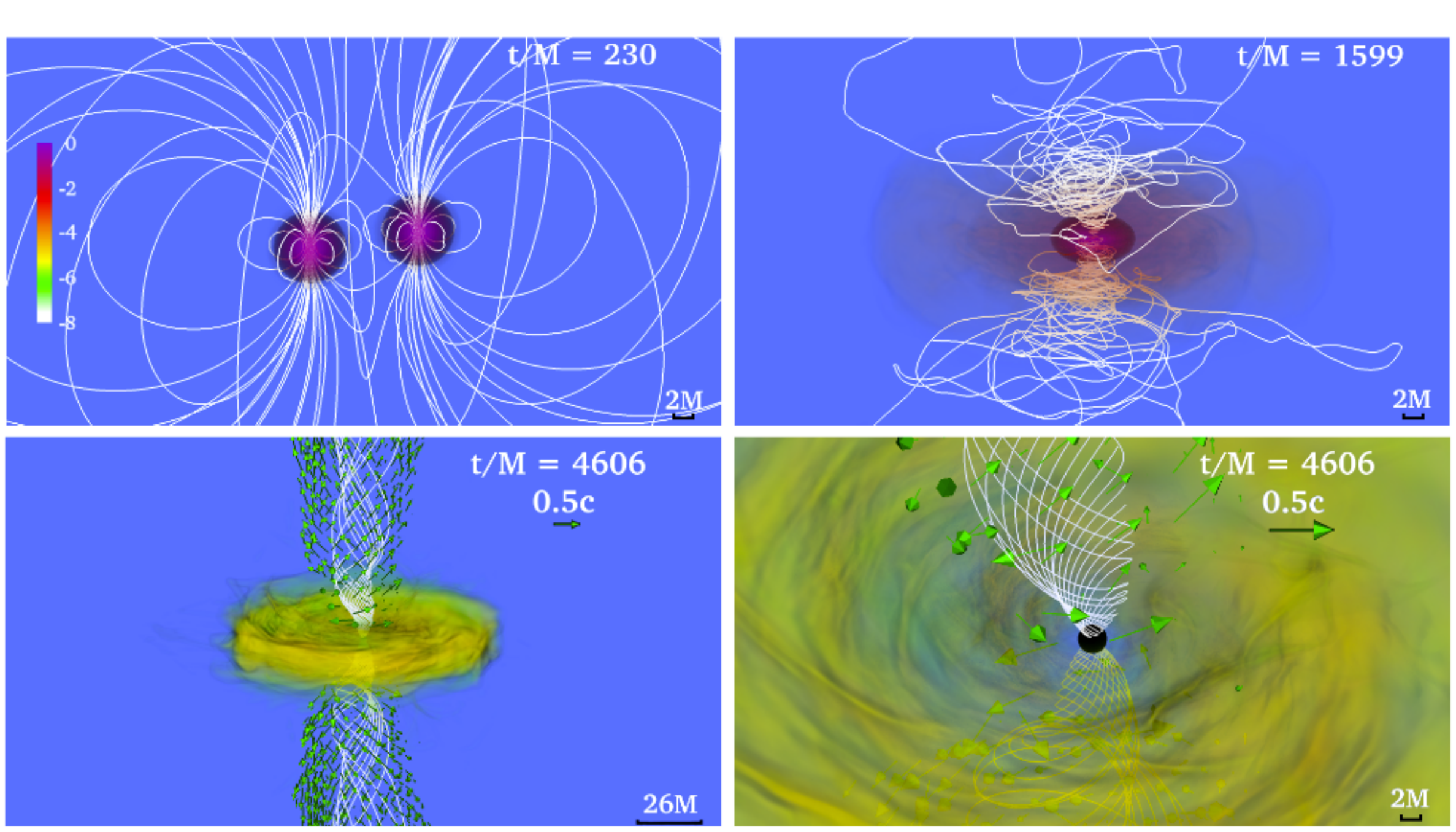}
\end{center}
\caption{GRMHD simulation of a BNS merger forming an accreting BH, in which an outflow emerges along the BH spin axis (adapted from \cite{Ruiz2016}).}
\label{fig4}       
\end{figure*}

A number of GRMHD simulations of BNS mergers addressed the problem of jet formation within the accreting BH central engine scenario 
(e.g., \cite{Rezzolla2011,Kiuchi2014,Dionysopoulou2015,Kawamura2016,Ruiz2016,Ruiz2020}).
These simulations confirmed the creation of a low density funnel along the BH spin axis of half-opening angle in the range $20^\circ\!-\!50^\circ$ and provided hints of an emerging magnetic field structure that could favour the formation of a SGRB jet. 
While no relativistic jet has been reported so far, the simulations presented in \cite{Ruiz2016} were the first to show the emergence of an outflow along the BH spin axis (Fig.~\ref{fig4}), following an analogous result obtained for NS--BH mergers \cite{Paschalidis2015}.
For the cases discussed in \cite{Ruiz2016}, the material in the funnel starts to acquire an outflowing velocity a few tens of ms after the collapse, when the magnetic-to-fluid energy density ratio inside the funnel has grown well above unity. The maximum Lorentz factor reported is $\Gamma\!\simeq\!1.25$, but the authors argue that terminal values up to $\sim\!100$ can in principle be achieved. 

Even though a demonstration that a SGRB jet can actually be produced in this context is still missing, results like the ones obtained in \cite{Ruiz2016} further reinforce the idea that the accreting BH scenario is a very promising possibility. 

\subsection{Massive neutron star central engine}
\label{magnetar}

A stable or metastable NS formed as the result of a BNS merger has a rotational energy of $\gtrsim\!10^{53}$\,erg and such a large energy reservoir, if efficiently channeled via the magnetic field, could in principle power a jet with properties consistent with a SGRB (e.g., \cite{Zhang2001,Gao2006,Metzger2008}). 
Indications that may favour this scenario come from the soft X-ray plateau features observed by Swift \cite{Gehrels2004-Swift} in a large fraction of SGRB events ($\gtrsim\!50\%$), following the prompt gamma-ray emission and lasting from minutes to hours (e.g., \cite{Rowlinson2010,Rowlinson2013,Lu2015}).  
The rather shallow decay of these signals appears inconsistent with the steeper lightcurves characterizing jet afterglows and suggests instead a persistent energy injection from a central source. While this emission may be difficult to explain in terms of a BH central engine, it could be naturally explained in presence of a long-lived magnetized NS (e.g., \cite{Rowlinson2010,Rowlinson2013,Yu2013,Lu2015,Metzger2014,Siegel2016a,Siegel2016b}; but see also \cite{Ciolfi2015,Ciolfi2018,Oganesyan2020} and refs.~therein).
The absence of a BH, on the other hand, could make the production of a relativistic outflow much more challenging. 
In particular, a main concern is that emergence of an incipient jet could be hampered by the much higher level of baryon pollution expected in this case along the remnant spin axis (e.g., \cite{Ciolfi2019}). 

GRMHD simulations based on idealized models of a differentially rotating massive NS (aimed at mimicking the outcome of a BNS merger) showed that an extended dipolar magnetic field aligned with the NS spin axis, in combination with the NS differential rotation, would produce an outflow \cite{Shibata2011,Kiuchi2012,Siegel2014,Ruiz2018}. Such an outflow is non-isotropic, with velocities that are higher at lower polar angles, resulting, at least in some cases, in a significant degree of collimation within a few hundred gravitational radii from the central compact object.
The outward acceleration, strongest along the spin axis, is the result of the growth of magnetic pressure associated with the winding up of the poloidal field lines, which acquire a helical structure. This result is obtained even with relatively low dipolar field strengths ($\sim\!10^{13}$\,G). The corresponding Poynting-flux luminosity scales with the square of the dipolar field strength and the luminosity of SGRBs can in principle be reproduced with $\sim\!10^{15}\!-\!10^{16}$\,G \cite{Shibata2011,Kiuchi2012,Siegel2014}.

The above findings, however, cannot be directly applied to a BNS merger because of two main limitations. 
First, an extended dipolar field is added by hand, while in actual BNS mergers the rather disordered initial magnetic field geometry of a NS remnant could take a very long time to acquire a large-scale helical structure along the spin axis (see Section \ref{geom}), which is a necessary condition to produce a collimated outflow \cite{Siegel2014}. 
Second, these simulations completely neglect the influence of the dense environment surrounding the bulk of the NS remnant (see Section \ref{remnant}), which opposes the emergence of a collimated outflow and may hamper it. 
As a consequence, while those simulations illustrate a potential jet launching mechanism, they cannot predict whether such a mechanism would be successful in more realistic conditions.

\begin{figure*}
\begin{center}
  \includegraphics[width=0.97\textwidth]{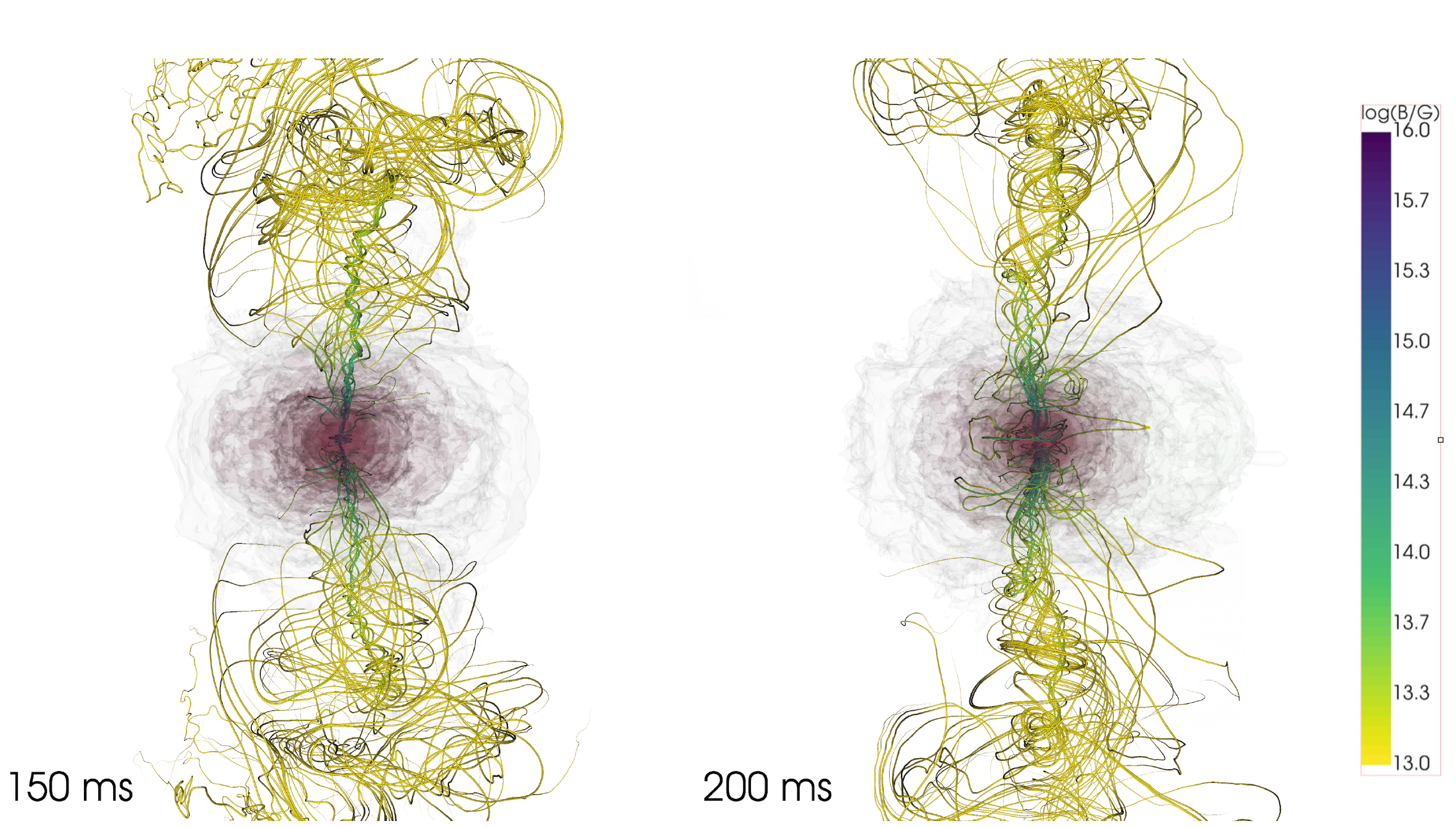}
\end{center}
\caption{Magnetic field structure 150 and 200\,ms after a BNS merger, with color-coded magnetic field strength (adapted from \cite{Ciolfi2020a}). The vertical scale covers $\pm\!2700$\,km along the remnant spin axis. Several semi-transperent isodensity surfaces are also shown for rest-mass densities between $6\!\times\!10^{6}$ and $4\!\times\!10^{9}$\,g/cm$^3$ (from light grey to red).}
\label{fig5}       
\end{figure*}

Most recently, a GRMHD BNS simulation with a long-lived NS remnant extending up to $\approx\!250$\,ms after merger (the longest to date) was presented in \cite{Ciolfi2020a} and, for the first time in a full merger simulation, the emergence of a magnetically-driven collimated outflow was reported (Fig.~\ref{fig5}). 
The simulation was long enough to show the early development of this outflow ($\approx\!100$\,ms after merger), its interaction with the isotropic baryon-loaded environment around the NS remnant (which also determines its collimation\footnote{The interaction between the outflow and the dense surrounding environment contributes to the differences in terms of collimation with respect to the findings of, e.g., \cite{Ruiz2018} (where the evolution also lasts over 200\,ms). }), its breaking out, and finally its gradual extinction.
This allowed for a detailed analysis, establishing that (i) the energy reservoir powering the outflow is given by differential rotation within the core of the NS remnant, as revealed by the evolution of outflow energy, rotational energy, and equatorial angular velocity profile, and (ii) the gradual development of a helical magnetic field structure along the remnant spin axis, sustained by differential rotation, is what provides the main acceleration. 
The above elements match the basic properties of the magnetorotational launching mechanism illustrated earlier in simulations of idealized differentially rotating NSs \cite{Shibata2011,Kiuchi2012,Siegel2014,Ruiz2018}. 
With this new result, we can conclude that a similar mechanism can actually operate in the context of BNS mergers. 

Besides demonstrating that the emergence of a collimated outflow from a massive NS remnant is possible, the results of \cite{Ciolfi2020a} also showed that such an outcome is not guaranteed for any highly magnetized NS remnant. 
Indeed, in a second simulation with a different initial magnetization, the production of a collimated outflow is hampered by the baryon-polluted environment surrounding the bulk of the NS remnant. 
This shows the critical role of the slow and nearly isotropic matter outflow characterizing magnetized NS remnants (see also Section \ref{remnant}).
Since the level of baryon-pollution around a NS remnant is also determined by the details of magnetic field amplification, it is not straightforward to predict whether a higher initial magnetization would always make the production of a collimated outflow easier. 

In order to support a scenario in which a SGRB jet is launched by a NS remnant rather than a BH, the collimated outflow found by \cite{Ciolfi2020a} would have to be able to evolve into a jet with specific properties, including a high enough jet core energy, collimation, and terminal Lorentz factor. 
On the contrary, the outflow properties are found largely inconsistent with what inferred for the jet of GRB\,170817A or any other known SGRB. 
While reconciling the total outflow energy of $\simeq\!3\times10^{49}$\,erg (maximum luminosity $\simeq\!5\times10^{50}$\,erg/s) and its collimation with the required values is challenging, but still conceivable, accelerating the outflow from Lorentz factors $\Gamma\!\lesssim\!1.05$ up to $\Gamma\!\gtrsim\!10$ is virtually impossible, as the outflow results at least three orders of magnitude too heavy to make this outcome attainable (energy-to-mass flux ratio is $<\!0.01$; also, magnetic energy is only $\sim\!23\%$ of the total energy).
In conclusion, for the case at hand, even if the NS remnant is able to produce a collimated outflow via the magnetorotational mechanism discussed above, it still has no chance to power a SGRB jet. 

The results of \cite{Ciolfi2020a} offer a rather strong case against the massive NS central engine scenario, pointing instead in favour of a BH origin for GRB\,170817A and SGRB in general. 
We stress, however, that such conclusion will have to be confirmed by exploring a larger variety of physical conditions (e.g., by including neutrino radiation and/or considering different EOS and mass ratios).\footnote{After the present review was submitted, Ref.~\cite{Moesta2020} further confirmed that superimposing by hand an extended dipolar magnetic field on a differentially rotating NS produces a collimated outflow, as found, e.g.,  in \cite{Shibata2011,Kiuchi2012,Siegel2014,Ruiz2018}.
In this case, the initial data were directly taken from the outcome of a nonmagnetized BNS merger simulation at 17\,ms after merger and neutrino radiation was included. While this represents an important step forward with respect to studies like \cite{Shibata2011,Kiuchi2012,Siegel2014,Ruiz2018}, a main caveat remains: there is no guarantee that collimated outflows with properties similar to those reported in \cite{Moesta2020}, which were obtained by imposing an ad hoc dipolar field at an arbitrary time, could also be obtained for magnetized BNSs going through the full merger process.}

\section{Mass ejection and kilonovae}
\label{ejectaKN}

BNS mergers are accompanied by a substantial mass ejection via different mechanisms, resulting in ejecta components with different dynamical properties and composition. 
For some of these ejecta components, the presence of magnetic fields can have a major impact on the outcome. 

The merger process itself is responsible for the so-called dynamical ejecta, including tidal and shock-driven ejecta. The former is induced by tidal forces during the last orbit before merger and the corresponding mass outflow is confined to the equatorial region. The latter is the result of the shock produced when the two NS cores crush into each other and of further shocks associated with the first, most violent oscillations of the merged object. In this case, the mass outflow is in all directions, although the contribution at large polar angles typically dominates over the contribution along the orbital axis. 
In both cases, the escaping velocity can be as high as $\sim\!0.3\,c$, while the total ejected mass is typically in the range $\sim\!10^{-3}\!-\!10^{-2}\,M_\odot$
(e.g., \cite{Hotokezaka2013,Bauswein2013,Radice2016,Sekiguchi2016}). The relative contribution of the two components depends on the binary total mass and mass ratio: higher total mass and mass ratio closer to unity tend to favour the shock-driven ejecta. 
For dynamical ejecta, magnetic field effects are typically negligible. 

Once the stable or metastable massive NS remnant has formed, material from the outer layers is pushed outwards by the growing pressure, which is enhanced by the increasing temperature, magnetic field amplification, (magneto)hydrodynamic turbulence, and neutrino heating (see also Section \ref{remnant}).
This results in a nearly isotropic outflow of matter, part of which can eventually become unbound and thus contribute to the merger ejecta (e.g., \cite{Ciolfi2017}). This outflow is typically rather slow close to the remnant ($<\!0.1~c$), but the ultimate velocity of the escaping material can be larger (see below). 

The presence of neutrino heating in the outer layers of the NS remnant can significantly enhance mass ejection in this phase (e.g., \cite{Dessart2009,Perego2014,Martin2015}). 
However, for the typical magnetizations expected after merger ($E_\mathrm{mag}\!\sim\!10^{50}\!-\!10^{51}$\,erg; see Section \ref{ampl}), the main driver of mass ejecton might be the magnetic field. 
For sufficiently long-lived NS remnants, the strong radial gradients of magnetic pressure and MHD turbulence can lead to cumulative mass outflows at 300\,km from the remnant as large as $\sim\!0.1\,M_\odot$  \cite{Ciolfi2017,Ciolfi2019}.
Even considering that only a fraction of this material is eventually becoming unbound, the total mass in this ejecta component may be comparable or even larger than the contribution from dynamical ejecta.

\begin{figure*}
\begin{center}
  \includegraphics[width=0.99\textwidth]{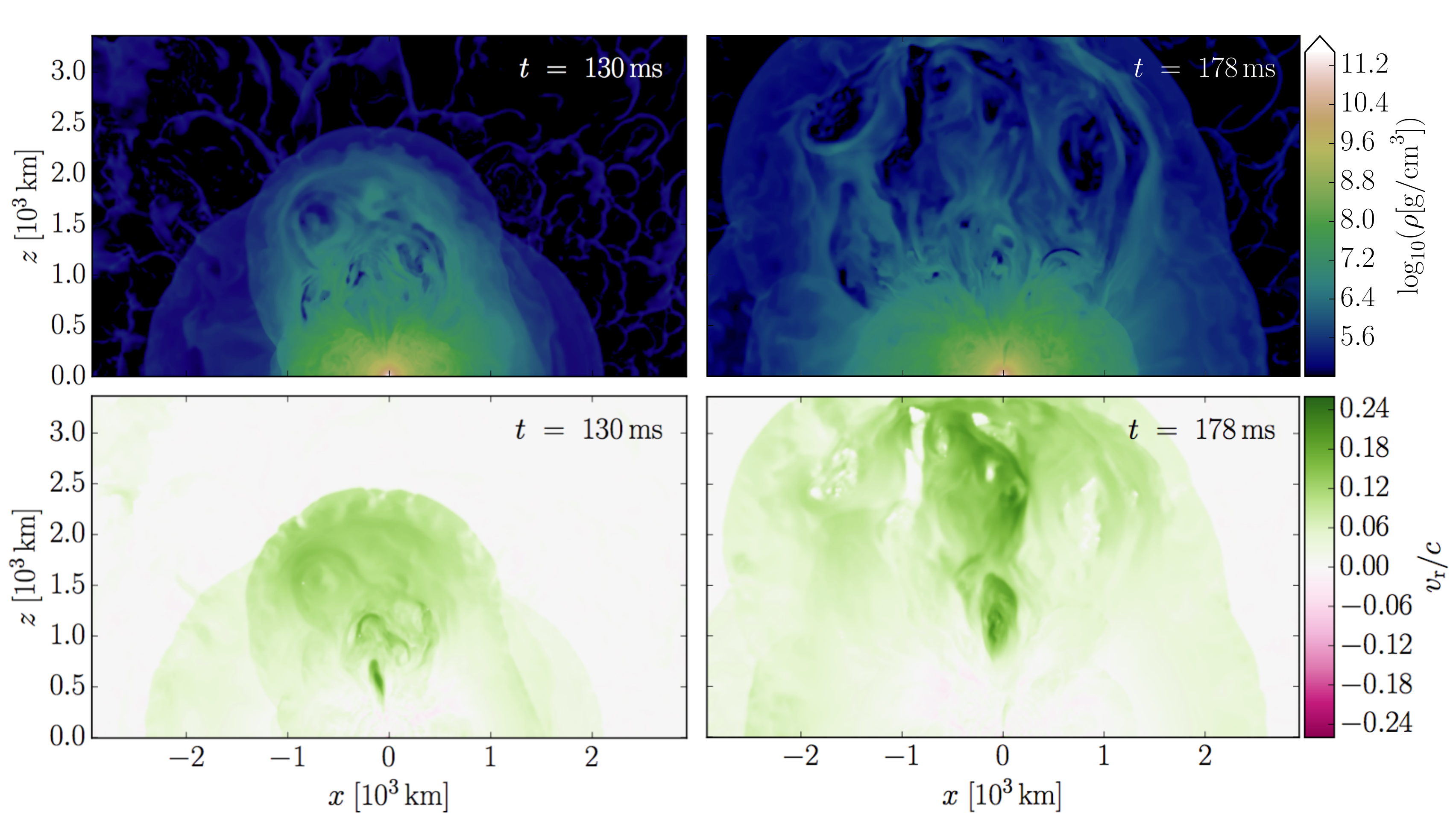}
\end{center}
\caption{Large scale meridional view of rest-mass density (top) and radial velocity (bottom) at 130 and 178\,ms after merger for a BNS merger simulation presented in \cite{Ciolfi2020a,CiolfiKalinani2020}.}
\label{fig6}       
\end{figure*}

As found in \cite{Ciolfi2020a,CiolfiKalinani2020}, if an ordered helical magnetic field structure is able to develop along the spin axis (see Sections \ref{magnetar}), mass ejection from the NS remnant will be dominated by a more collimated outflow component, mostly contained within an half-opening angle of $\simeq\!15^\circ$ from the axis. In this case, the radial velocity of the material can become as large as $\simeq\!0.2\,c$ (Fig.~\ref{fig6}).
As a note of caution, the fraction of BNS mergers resulting in such a collimated outflow is unknown and when this is not the case, the typical outflow velocities at large distance are more likely to remain limited to $\sim\!0.1\,c$ \cite{Ciolfi2020a}.
We also note that the above indications are based on simulations that do not include neutrino radiation, nor the possibility of nuclear recombination. The latter effects, if dynamically relevant, could further increase the fraction of material that eventually becomes unbound as well as the final escaping velocity. 

After the eventual collapse to a BH (if any), the hot and thick accretion disk is responsible for additional mass ejection (e.g, \cite{Fernandez2013,Just2015,Siegel2017,Siegel2018,Fernandez2019,Fujibayashi2020}). 
The MHD turbulence in the disk and neutrino heating contribute to launch a slow and persistent matter outflow, which can in the long term (i.e. on viscous timescales) unbind up to $\sim\!40\%$ of the disk mass, corresponding to $\sim\!0.01-0.1\,M_\odot$ (e.g, \cite{Siegel2017,Siegel2018,Fernandez2019}). 
In terms of mass, this ejecta component likely represents the dominant one in most BNS mergers. 
Also in this case, the presence of magnetic fields can significantly affect the amount of ejecta. Recent simulations showed that the ejection of mass can increase by a factor of $\sim\!2$ compared to nonmagnetized cases (e.g, \cite{Fernandez2019}). 
Mass ejection from the accretion disk is highly isotropic and the typical asymptotic escape velocity is $\sim\!0.1\,c$.

\subsection{AT\,2017gfo: the kilonova of August 2017}
\label{KN2017}

Along with the first GW signal from a BNS merger and the association with a SGRB and the accompanying multiwavelength afterglows, the multimessenger discovery of August 2017 also led to the identification of an optical/infrared signal fully consistent with a so-called ``kilonova'', a thermal transient powered by the radioactive decay of heavy nuclei synthesized via r-process nucleosynthesis within the material ejected by the merger (e.g., \cite{Metzger2019LRR} and refs.~therein).

After a BNS merger, the composition and thermodynamical history of each fluid element within the ejected material determines how far the r-process nucleosynthesis can advance in producing heavier and heavier elements (e.g., \cite{Lippuner2015}). In turn, the nucleosynthesis products determine the opacity of the material as well as the radioactive heating that will ultimately power the kilonova (e.g., \cite{Barnes2013}). 
The electron fraction $Y_e$ of a fluid element is a crucial parameter. Neutron-rich material ($Y_e\!\lesssim\!0.25$) can synthesize up very heavy r-process elements (i.e.~up to atomic mass numbers $A\!>\!140$), including the group of lanthanides, and the resulting opacity of $\sim\!10$\,cm$^2$/g is much higher than that of iron-group elements \cite{Kasen2013,Tanaka2013}. Neutron-poor ejecta ($Y_e\!\gtrsim\!0.25$), on the other hand, only produce elements up to atomic mass numbers $A\!\lesssim\!140$ (i.e.~up to light, second-peak r-process nuclei), which keeps the opacity lower ($\sim\!0.1\!-\!1$\,cm$^2$/g).

Assuming a single, uniform, and isotropic ejecta component, its mass, characteristic velocity, and opacity are the fundamental parameters to establish the peak time and peak luminosity of the resulting kilonova transient, as well as the corresponding effective temperature (e.g., \cite{Grossman2014}).
A real kilonova, however, can be the result of a complex combination of ejecta components with diverse composition and geometrical distribution, where the velocity and opacity may also have a significant dependence on the polar angle. 
Moreover, the opacity is in general wavelength-dependent and a function of temperature, implying that it can also vary in time.
Besides the intrinsic complexity of the problem and the uncertainties on various microphysical parameters (opacities, heating rates, thermalization efficiencies, and more), current models are also limited by the fact that proper radiation transport simulations are too computationally expensive, forcing the use of approximations (e.g., \cite{Metzger2019LRR} and refs.~therein).

Despite the above difficulties, a number of authors attempted to model the kilonova transient of August 2017, named AT\,2017gfo. 
A first key result, common to the different models, is that a single component with 
roughly constant and uniform opacity seems to be unable to fit the data. The transient is instead reproduced with at least two distinct components (e.g., \cite{Kasen2017}; see also \cite{Metzger2019LRR} and refs.~therein): (i) a fast evolving ``blue'' component with mass $\approx\!1.5\!-\!2.5\!\times\!10^{-2}\,M_\odot$, velocity $\approx\!0.2\!-\!0.3\,c$, and relatively low opacity of $\approx\!0.5$\,cm$^2$/g (suggesting lanthanide-free material, with $Y_e\!\gtrsim\!0.25$), peaking at $\sim\!1$\,day after merger, and (ii) a lanthanide-rich ``red'' component with mass $\approx\!4\!-\!6\!\times\!10^{-2}\,M_\odot$, velocity $\approx\!0.1\,c$, and a much higher opacity of $\sim\!10$\,cm$^2$/g, emerging on a longer timescale of $\sim\!1$\,week.
Models with three components have also been proposed (e.g., \cite{Perego2017b,Villar2017}).

The association of the blue and red kilonova components with specific types of merger ejecta is not obvious and different interpretations exist. 
The red component appears too massive and too slow to be explained via dynamical ejecta, while various authors showed that baryon-loaded winds launched by the accretion disk after BH formation represent a viable explanation (e.g., \cite{Siegel2018}). 
Mass ejection from the NS remnant (i.e.~prior to collapse) would instead hardly match the lanthanide-rich requirement. The copious neutrino radiation passing through this material is indeed expected to significantly raise its electron fraction, strongly limiting the production of lanthanides and leading to a blue type of kilonova (e.g., \cite{Perego2014}).

The latter consideration, on the other hand, makes this component a possible candidate to explain the more puzzling blue part of AT\,2017gfo, which is at the same time rather fast and rather massive. 
The presence of strong magnetic fields, enhancing mass ejection and possibly able to accelerate the material up to the required asymptotic velocity (see Fig.~\ref{fig6}), represents a crucial ingredient to support this interpretation \cite{Metzger2018}.
Recent results from \cite{CiolfiKalinani2020} showed that the magnetically driven baryon wind from the NS remnant represents indeed a viable explanation for the blue kilonova, being potentially able to reproduce the mass, velocity, and opacity inferred from the AT\,2017gfo observations.
Among the proposed alternatives, we mention here the model presented in \cite{Kawaguchi2018}, where mass ejection is mainly attributed to the nearly isotropic outflow from the NS remnant, while the transition from blue to red kilonova and from fast to slow photospheric velocity is explained via the reprocessing of the emitted radiation across the surrounding dynamical ejecta, assuming a specific angular distribution. We also mention a possible explanation of the blue kilonova component based on dynamical mass ejection from the development of spiral arms in the newly formed NS remnant \cite{Nedora2019}.
We refer the reader to \cite{Shibata2019,Metzger2019LRR} and refs.~therein for a more comprehensive review of the variety of models present in the literature. 
\begin{figure*}
\begin{center}
  \includegraphics[width=0.97\textwidth]{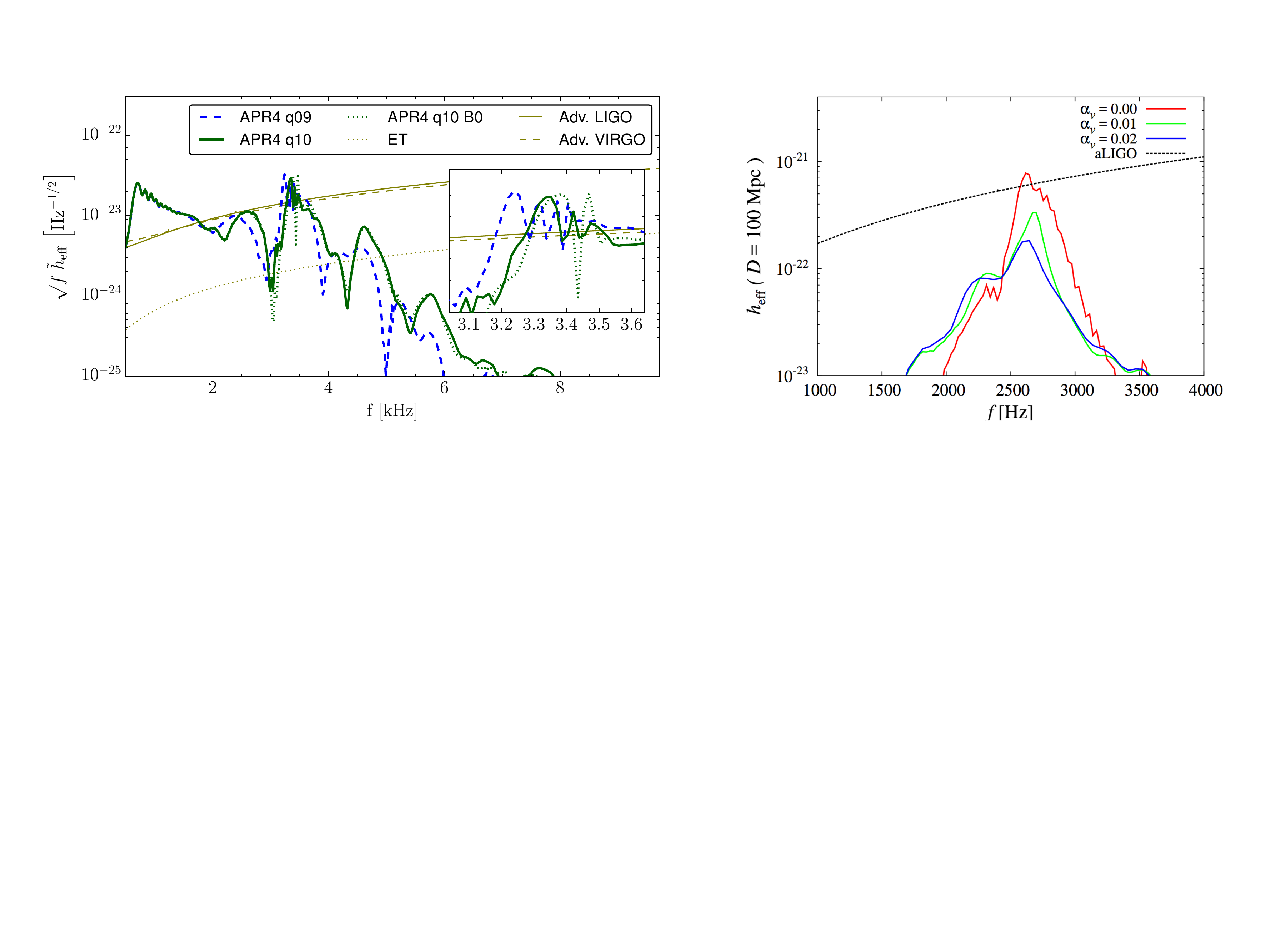}
\end{center}
\caption{
{\it Left}\,(a): Power spectrum of the GW strain assuming 100\,Mpc distance for different BNS merger models presented in \cite{Ciolfi2017}, compared to the design sensitivity curves of current (Advanced LIGO/Virgo) and planned (Einstein Telescope) gravitational wave detectors. Note that the model labelled ``q10 B0'' is the nonmagnetized equivalent of the ``q10'' model. {\it Right}\,(b): Post-merger GW signal for a BNS merger model with varying effective viscosity parameter (adapted from \cite{Shibata2017b}).}
\label{fig7}       
\end{figure*}

\section{Gravitational wave emission}
\label{gw}

The GW signal form a BNS merger can be separated into inspiral, merger, and post-merger phases, where the central merger phase refers to a window of a few ms around the time when the two NS cores touch each other or, conventionally, the time of maximum amplitude of the GW signal. 
Magnetic fields are expected to have negligible effects on the GW emission during the inspiral phase up to merger \cite{Giacomazzo2009}, but they can be relevant for the post-merger signal. 
The emission from the stable or metastable massive NS remnant, which is absent in the case of prompt collapse to a BH, is dominated by a single frequency, corresponding to a fundamental $l\!=\!m\!=\!2$ oscillation mode (e.g., \cite{Bauswein2012}; see also, e.g., \cite{Paschalidis2017LRR} and refs.~therein). 
This characteristic frequency has a strong correlation with the NS EOS (e.g., \cite{Stergioulas2011,Bauswein2012,Hotokezaka2013b,Takami2014,Bernuzzi2015,Baiotti2017,Paschalidis2017LRR}) and therefore a secure detection of this part of the GW signal would offer a unique opportunity to improve present constraints on the behaviour of matter at supranuclear densities (e.g., \cite{Clark2014,Yang2018}). 

Early after merger, magnetic fields are strongly amplified until they start becoming dynamically relevant (See Section \ref{ampl}). This process affects the redistribution of angular momentum and can significantly alter the GW signal. The characteristic frequency can be shifted (typically to lower values; see Fig.~\ref{fig7}a), but such modification is most likely not substantial enough to be observable in realistic cases of post-merger GW signal detections, even for very high magnetizations \cite{Ciolfi2017}. 
The development of MHD turbulence and the resulting effective viscosity, on the other hand, may have a detectable influence on the signal amplitude. Besides accelerating the redistribution of angular momentum, the MHD effective viscosity could indeed considerably damp the NS oscillation modes. As a result, the amplitude of the post-merger signal would be reduced \cite{Shibata2017b} (Fig.~\ref{fig7}b).
Nevertheless, whether realistic values of the effective viscosity are sufficient to lead to an appreciable damping is still matter of debate (e.g., \cite{Radice2017}).

\section{Summary and outlook}
\label{concl}

The multimessenger discovery of August 2017 confirmed the major importance of BNS mergers for astrophysics and fundamental physics. 
Along with this breakthrough, the long-awaited evidence for the connection with SGRBs has been obtained, as well as strong indications that these systems are ideal sites for the nucleosynthesis of heavy r-process elements.
Despite the extraordinary step forward, however, there is still a lot to be understood on the merger process and further investigation is necessary before we can establish a firm link between the specific properties of a merging BNS and the corresponding EM and GW emission. 

In this review, we reported on the status of theoretical modelling of BNS mergers based on numerical relativity simulations, focussing the attention on the magnetic field as a key ingredient. 
First, we discussed the evolution and amplification of magnetic fields during and after a BNS merger.
Thanks to the substantial progress of GRMHD simulations, there is today a general consensus that a strong amplification up to a physical saturation is most likely to produce magnetization levels that are dynamically important, with effects on the properties of the (meta)stable NS remnant and the surrounding environment.
Moreover, recent simulations provided hints of an emerging large-scale magnetic field structure, supporting the idea that a massive NS remnant, as long as no collapse to a BH occurs, could build up a global dipolar magnetic field of magnetar-like strength. 
On the other hand, present studies are still affected by various limitations.
A critical one is that the resolution currently affordable is known to be insufficient to fully capture the dominant magnetic field amplification mechanisms (in particular the KH instability).
 
The pivotal role of magnetic fields in the production of SGRB jets is now well established, but important doubts remain on the involved physical processes. 
The most promising scenario is the one based on an accreting BH central engine. GRMHD simulations in this context provided encouraging indications, even though the production of a jet with properties consistent with a SGRB is yet to be demonstrated. Moreover, is not yet clear whether the jet would be launched via the Blandford--Znajek mechanism or by other means. The alternative SGRB scenario in which the central engine is a massive NS remnant appears at this time less favourable, mainly due to the much higher level of baryon pollution opposing the emergence of an incipient jet.
Recent simulations of this scenario showed that the production of a collimated magnetically-driven outflow is possible, although not necessarily ubiquitous.
Neverhteless, the outflow properties and in particular the very limited terminal Lorentz factor are found largely inconsistent with a SGRB jet. 
This result points in disfavour of a NS central engine. We caution, however, that a larger variety of physical conditions needs to be explored in order to reach a final conclusion. 
The issue on the nature of the central engine also relates to the interpretation of soft X-ray emission features accompanying a large fraction of SGRBs, in particular the so-called X-ray plateaus. 
If the spindown radiation from a long-lived magnetized NS is to be excluded as the source powering X-ray plateaus, an alternative explanation needs to be found that is compatible with an accreting BH central engine. 

Another key question left behind by the August 2017 event concerns the ejecta components that originated the kilonova transient. 
Magnetic fields certainly represent an important element in this open problem, given their impact on post-merger mass ejection both from the massive NS remnant and, after the eventual collapse to a BH, from the accretion disk.
In addition, the rather massive, fast, and lanthanide-poor ejecta component necessary to explain the early ``blue'' kilonova might be explained in terms of a magnetically-driven outflow from the NS remnant. If confirmed, this would further identify the magnetic field as a crucial ingredient that needs to be included for a proper modelling of kilonovae. 

We concluded our review by discussing the influence of magnetic fields on the post-merger GW signal. High magnetization levels can alter the characteristic frequency of the emission, although the corresponding effect is likely not large enough to be directly detectable. 
On the other hand, the effective viscosity induced by MHD turbulence might significantly reduce the signal amplitude, lowering the prospects of a future detection. 
Nevertheless, the extent of such amplitude damping remains matter of debate and it is not clear whether this could actually have an impact. 

As the prospects of multimessenger BNS merger detections will keep increasing in the coming years, the growing theoretical effort to understand the physics of these systems will become more urgent and, probably, more challenging. 
At the same time, the progress of models and numerical simulations, aided by the parallel advancement of computational resources, will give us a unique opportunity to exploit the richness of information encoded in the GW and EM signals. 
In this context, magnetic fields will likely remain an indispensable ingredient for any meaningful model attempting an interpretation of the new observations.

\begin{acknowledgements}

We thank Daniele Vigan\`o, David Radice, Bruno Giacomazzo, Wolfgang Kastaun, Davide Lazzati, and Albino Perego for useful discussions. 

\end{acknowledgements}


\begin{thebibliography}{100}
\providecommand{\url}[1]{{#1}}
\providecommand{\urlprefix}{URL }
\expandafter\ifx\csname urlstyle\endcsname\relax
  \providecommand{\doi}[1]{DOI~\discretionary{}{}{}#1}\else
  \providecommand{\doi}{DOI~\discretionary{}{}{}\begingroup
  \urlstyle{rm}\Url}\fi

\bibitem{LVC-Hubble}
{Abbott}, B.P., {Abbott}, R., {Abbott}, T.D., {Acernese}, F., {Ackley}, K.,
  {Adams}, C., {Adams}, T., {Addesso}, P., {Adhikari}, R.X., {Adya}, V.B.,
  et~al.: {A gravitational-wave standard siren measurement of the Hubble
  constant}.
\newblock Nature \textbf{551}, 85 (2017).
\newblock \doi{10.1038/nature24471}

\bibitem{LVC-GRB}
{Abbott}, B.P., {Abbott}, R., {Abbott}, T.D., {Acernese}, F., {Ackley}, K.,
  {Adams}, C., {Adams}, T., {Addesso}, P., {Adhikari}, R.X., {Adya}, V.B.,
  et~al.: {Gravitational Waves and Gamma-Rays from a Binary Neutron Star
  Merger: GW170817 and GRB 170817A}.
\newblock Astrophys. J. Lett. \textbf{848}, L13 (2017).
\newblock \doi{10.3847/2041-8213/aa920c}

\bibitem{LVC-BNS}
{Abbott}, B.P., {Abbott}, R., {Abbott}, T.D., {Acernese}, F., {Ackley}, K.,
  {Adams}, C., {Adams}, T., {Addesso}, P., {Adhikari}, R.X., {Adya}, V.B.,
  et~al.: {GW170817: Observation of Gravitational Waves from a Binary Neutron
  Star Inspiral}.
\newblock Phys. Rev. Lett. \textbf{119}(16), 161101 (2017).
\newblock \doi{10.1103/PhysRevLett.119.161101}

\bibitem{LVC-MMA}
{Abbott}, B.P., {Abbott}, R., {Abbott}, T.D., {Acernese}, F., {Ackley}, K.,
  {Adams}, C., {Adams}, T., {Addesso}, P., {Adhikari}, R.X., {Adya}, V.B.,
  et~al.: {Multi-messenger Observations of a Binary Neutron Star Merger}.
\newblock Astrophys. J. Lett. \textbf{848}, L12 (2017).
\newblock \doi{10.3847/2041-8213/aa91c9}

\bibitem{LVC-170817properties}
{Abbott}, B.P., {Abbott}, R., {Abbott}, T.D., {Acernese}, F., {Ackley}, K.,
  {Adams}, C., {Adams}, T., {Addesso}, P., {Adhikari}, R.X., {Adya}, V.B.,
  et~al.: {Properties of the Binary Neutron Star Merger GW170817}.
\newblock Phys. Rev. X \textbf{9}(1), 011001 (2019).
\newblock \doi{10.1103/PhysRevX.9.011001}

\bibitem{Alexander2017}
{Alexander}, K.D., {Berger}, E., {Fong}, W., {Williams}, P.K.G., {Guidorzi},
  C., {Margutti}, R., {Metzger}, B.D., {Annis}, J., {Blanchard}, P.K., {Brout},
  D., et~al.: {The
  Electromagnetic Counterpart of the Binary Neutron Star Merger LIGO/Virgo
  GW170817. VI. Radio Constraints on a Relativistic Jet and Predictions for
  Late-time Emission from the Kilonova Ejecta}.
\newblock Astrophys. J. Lett. \textbf{848}, L21 (2017).
\newblock \doi{10.3847/2041-8213/aa905d}

\bibitem{Alexander2018}
{Alexander}, K.D., {Margutti}, R., {Blanchard}, P.K., {Fong}, W., {Berger}, E.,
  {Hajela}, A., {Eftekhari}, T., {Chornock}, R., {Cowperthwaite}, P.S.,
  {Giannios}, D., et~al.: {A Decline in the X-Ray through Radio
  Emission from GW170817 Continues to Support an Off-axis Structured Jet}.
\newblock Astrophys. J. Lett. \textbf{863}, L18 (2018).
\newblock \doi{10.3847/2041-8213/aad637}

\bibitem{Aloy2005}
{Aloy}, M.A., {Janka}, H.T., {M{\"u}ller}, E.: {Relativistic outflows from
  remnants of compact object mergers and their viability for short gamma-ray
  bursts}.
\newblock Astron. Astrophys. \textbf{436}(1), 273 (2005).
\newblock \doi{10.1051/0004-6361:20041865}

\bibitem{Anderson2008}
{Anderson}, M., {Hirschmann}, E.W., {Lehner}, L., {Liebling}, S.L., {Motl},
  P.M., {Neilsen}, D., {Palenzuela}, C., {Tohline}, J.E.: {Magnetized
  Neutron-Star Mergers and Gravitational-Wave Signals}.
\newblock Phys. Rev. Lett. \textbf{100}(19), 191101 (2008).
\newblock \doi{10.1103/PhysRevLett.100.191101}

\bibitem{Arcavi2017}
{Arcavi}, I., {Hosseinzadeh}, G., {Howell}, D.A., {McCully}, C., {Poznanski},
  D., {Kasen}, D., {Barnes}, J., {Zaltzman}, M., {Vasylyev}, S., {Maoz}, D., et~al.: {Optical emission from a kilonova following a
  gravitational-wave-detected neutron-star merger}.
\newblock Nature \textbf{551}(7678), 64 (2017).
\newblock \doi{10.1038/nature24291}

\bibitem{Baiotti2017}
{Baiotti}, L., {Rezzolla}, L.: {Binary neutron star mergers: a review of
  Einstein{\textquoteright}s richest laboratory}.
\newblock Rep. Prog. Phys. \textbf{80}(9), 096901 (2017).
\newblock \doi{10.1088/1361-6633/aa67bb}

\bibitem{Balbus1991}
{Balbus}, S.A., {Hawley}, J.F.: {A powerful local shear instability in weakly
  magnetized disks. I - Linear analysis. II - Nonlinear evolution}.
\newblock Astrophys. J. \textbf{376}, 214 (1991).
\newblock \doi{10.1086/170270}

\bibitem{Barnes2013}
{Barnes}, J., {Kasen}, D.: {Effect of a High Opacity on the Light Curves of
  Radioactively Powered Transients from Compact Object Mergers}.
\newblock Astrophys. J. \textbf{775}(1), 18 (2013).
\newblock \doi{10.1088/0004-637X/775/1/18}

\bibitem{Barthelmy2005a}
{Barthelmy}, S.D., {Chincarini}, G., {Burrows}, D.N., {Gehrels}, N., {Covino},
  S., {Moretti}, A., {Romano}, P., {O'Brien}, P.T., {Sarazin}, C.L.,
  {Kouveliotou}, C., et~al.: {An origin for short {$\gamma$}-ray
  bursts unassociated with current star formation}.
\newblock Nature \textbf{438}, 994 (2005).
\newblock \doi{10.1038/nature04392}

\bibitem{Bauswein2013}
{Bauswein}, A., {Goriely}, S., {Janka}, H.T.: {Systematics of Dynamical Mass
  Ejection, Nucleosynthesis, and Radioactively Powered Electromagnetic Signals
  from Neutron-star Mergers}.
\newblock Astrophys. J. \textbf{773}(1), 78 (2013).
\newblock \doi{10.1088/0004-637X/773/1/78}

\bibitem{Bauswein2012}
{Bauswein}, A., {Janka}, H.T.: {Measuring Neutron-Star Properties via
  Gravitational Waves from Neutron-Star Mergers}.
\newblock Phys. Rev. Lett. \textbf{108}(1), 011101 (2012).
\newblock \doi{10.1103/PhysRevLett.108.011101}

\bibitem{Berger2014}
{Berger}, E.: {Short-Duration Gamma-Ray Bursts}.
\newblock Ann. Rev. Astron. Astrophys. \textbf{52}, 43 (2014).
\newblock \doi{10.1146/annurev-astro-081913-035926}

\bibitem{Berger2013}
{Berger}, E., {Fong}, W., {Chornock}, R.: {An r-process Kilonova Associated
  with the Short-hard GRB 130603B}.
\newblock Astrophys. J. Lett. \textbf{774}, L23 (2013).
\newblock \doi{10.1088/2041-8205/774/2/L23}

\bibitem{Bernuzzi2015}
{Bernuzzi}, S., {Dietrich}, T., {Nagar}, A.: {Modeling the Complete
  Gravitational Wave Spectrum of Neutron Star Mergers}.
\newblock Phys. Rev. Lett. \textbf{115}(9), 091101 (2015).
\newblock \doi{10.1103/PhysRevLett.115.091101}

\bibitem{Bilous2019}
{Bilous}, A.V., {Watts}, A.L., {Harding}, A.K., {Riley}, T.E., {Arzoumanian},
  Z., {Bogdanov}, S., {Gendreau}, K.C., {Ray}, P.S., {Guillot}, S., {Ho},
  W.C.G., et~al.: {A NICER View of PSR J0030+0451: Evidence for a
  Global-scale Multipolar Magnetic Field}.
\newblock Astrophys. J. Lett. \textbf{887}(1), L23 (2019).
\newblock \doi{10.3847/2041-8213/ab53e7}

\bibitem{Blandford1977}
{Blandford}, R.D., {Znajek}, R.L.: {Electromagnetic extraction of energy from
  Kerr black holes}.
\newblock Mon. Not. R. Astron. Soc. \textbf{179}, 433 (1977).
\newblock \doi{10.1093/mnras/179.3.433}

\bibitem{Carrasco2020}
{Carrasco}, F., {Vigan{\`o}}, D., {Palenzuela}, C.: {Gradient subgrid-scale
  model for relativistic MHD large-eddy simulations}.
\newblock Phys. Rev. D \textbf{101}(6), 063003 (2020).
\newblock \doi{10.1103/PhysRevD.101.063003}

\bibitem{Ciolfi2018}
{Ciolfi}, R.: {Short gamma-ray burst central engines}.
\newblock Int. J. Mod. Phys. D \textbf{27}, 1842004 (2018).
\newblock \doi{10.1142/S021827181842004X}

\bibitem{Ciolfi2020c}
{Ciolfi}, R.: {Binary neutron star mergers after GW170817}.
\newblock arXiv e-prints arXiv:2005.02964 (2020)

\bibitem{Ciolfi2020a}
{Ciolfi}, R.: {Collimated outflows from long-lived binary neutron star merger
  remnants}.
\newblock Mon. Not. R. Astron. Soc. Lett. \textbf{495}(1), L66 (2020).
\newblock \doi{10.1093/mnrasl/slaa062}

\bibitem{CiolfiKalinani2020}
{Ciolfi}, R., {Kalinani}, J.V.: {Magnetically driven baryon winds from binary
  neutron star merger remnants and the blue kilonova of August 2017}.
\newblock arXiv e-prints arXiv:2004.11298 (2020)

\bibitem{Ciolfi2017}
{Ciolfi}, R., {Kastaun}, W., {Giacomazzo}, B., {Endrizzi}, A., {Siegel}, D.M.,
  {Perna}, R.: {General relativistic magnetohydrodynamic simulations of binary
  neutron star mergers forming a long-lived neutron star}.
\newblock Phys. Rev. D \textbf{95}(6), 063016 (2017).
\newblock \doi{10.1103/PhysRevD.95.063016}

\bibitem{Ciolfi2019}
{Ciolfi}, R., {Kastaun}, W., {Kalinani}, J.V., {Giacomazzo}, B.: {First 100 ms
  of a long-lived magnetized neutron star formed in a binary neutron star
  merger}.
\newblock Phys. Rev. D \textbf{100}(2), 023005 (2019).
\newblock \doi{10.1103/PhysRevD.100.023005}

\bibitem{Ciolfi2013}
{Ciolfi}, R., {Rezzolla}, L.: {Twisted-torus configurations with large toroidal
  magnetic fields in relativistic stars.}
\newblock Mon. Not. R. Astron. Soc. Lett. \textbf{435}, L43 (2013).
\newblock \doi{10.1093/mnrasl/slt092}

\bibitem{Ciolfi2015}
{Ciolfi}, R., {Siegel}, D.M.: {Short Gamma-Ray Bursts in the ``Time-reversal''
  Scenario}.
\newblock Astrophys. J. Lett. \textbf{798}, L36 (2015).
\newblock \doi{10.1088/2041-8205/798/2/L36}

\bibitem{Clark2014}
{Clark}, J., {Bauswein}, A., {Cadonati}, L., {Janka}, H.T., {Pankow}, C.,
  {Stergioulas}, N.: {Prospects for high frequency burst searches following
  binary neutron star coalescence with advanced gravitational wave detectors}.
\newblock Phys. Rev. D \textbf{90}(6), 062004 (2014).
\newblock \doi{10.1103/PhysRevD.90.062004}

\bibitem{Coulter2017}
{Coulter}, D.A., {Foley}, R.J., {Kilpatrick}, C.D., {Drout}, M.R., {Piro},
  A.L., {Shappee}, B.J., {Siebert}, M.R., {Simon}, J.D., {Ulloa}, N., {Kasen},
  D., et~al.: {Swope Supernova Survey
  2017a (SSS17a), the optical counterpart to a gravitational wave source}.
\newblock Science \textbf{358}(6370), 1556 (2017).
\newblock \doi{10.1126/science.aap9811}

\bibitem{Dessart2009}
{Dessart}, L., {Ott}, C.D., {Burrows}, A., {Rosswog}, S., {Livne}, E.:
  {Neutrino Signatures and the Neutrino-Driven Wind in Binary Neutron Star
  Mergers}.
\newblock Astrophys. J. \textbf{690}(2), 1681 (2009).
\newblock \doi{10.1088/0004-637X/690/2/1681}

\bibitem{Dionysopoulou2015}
{Dionysopoulou}, K., {Alic}, D., {Rezzolla}, L.: {General-relativistic
  resistive-magnetohydrodynamic simulations of binary neutron stars}.
\newblock Phys. Rev. D \textbf{92}(8), 084064 (2015).
\newblock \doi{10.1103/PhysRevD.92.084064}

\bibitem{Duez2006a}
{Duez}, M.D., {Liu}, Y.T., {Shapiro}, S.L., {Shibata}, M., {Stephens}, B.C.:
  {Collapse of Magnetized Hypermassive Neutron Stars in General Relativity}.
\newblock Phys. Rev. Lett. \textbf{96}(3), 031101 (2006).
\newblock \doi{10.1103/PhysRevLett.96.031101}

\bibitem{Duez2004}
{Duez}, M.D., {Liu}, Y.T., {Shapiro}, S.L., {Stephens}, B.C.: {General
  relativistic hydrodynamics with viscosity: Contraction, catastrophic
  collapse, and disk formation in hypermassive neutron stars}.
\newblock Phys. Rev. D \textbf{69}(10), 104030 (2004).
\newblock \doi{10.1103/PhysRevD.69.104030}

\bibitem{Duez2019}
{Duez}, M.D., {Zlochower}, Y.: {Numerical relativity of compact binaries in the
  21st century}.
\newblock Rep. Prog. Phys. \textbf{82}(1), 016902 (2019).
\newblock \doi{10.1088/1361-6633/aadb16}

\bibitem{East2019}
{East}, W.E., {Paschalidis}, V., {Pretorius}, F., {Tsokaros}, A.: {Binary
  neutron star mergers: Effects of spin and post-merger dynamics}.
\newblock Phys. Rev. D \textbf{100}(12), 124042 (2019).
\newblock \doi{10.1103/PhysRevD.100.124042}

\bibitem{Eichler1989}
{Eichler}, D., {Livio}, M., {Piran}, T., {Schramm}, D.N.: {Nucleosynthesis,
  neutrino bursts and gamma-rays from coalescing neutron stars}.
\newblock Nature \textbf{340}, 126 (1989).
\newblock \doi{10.1038/340126a0}

\bibitem{Endrizzi2016}
{Endrizzi}, A., {Ciolfi}, R., {Giacomazzo}, B., {Kastaun}, W., {Kawamura}, T.:
  {General relativistic magnetohydrodynamic simulations of binary neutron star
  mergers with the APR4 equation of state}.
\newblock Class. Quantum Grav. \textbf{33}(16), 164001 (2016).
\newblock \doi{10.1088/0264-9381/33/16/164001}

\bibitem{Faber2012}
{Faber}, J.A., {Rasio}, F.A.: {Binary Neutron Star Mergers}.
\newblock Liv. Rev. Rel. \textbf{15}(1), 8 (2012).
\newblock \doi{10.12942/lrr-2012-8}

\bibitem{Fernandez2013}
{Fern{\'a}ndez}, R., {Metzger}, B.D.: {Delayed outflows from black hole
  accretion tori following neutron star binary coalescence}.
\newblock Mon. Not. R. Astron. Soc. \textbf{435}(1), 502 (2013).
\newblock \doi{10.1093/mnras/stt1312}

\bibitem{Fernandez2019}
{Fern{\'a}ndez}, R., {Tchekhovskoy}, A., {Quataert}, E., {Foucart}, F.,
  {Kasen}, D.: {Long-term GRMHD simulations of neutron star merger accretion
  discs: implications for electromagnetic counterparts}.
\newblock Mon. Not. R. Astron. Soc. \textbf{482}(3), 3373 (2019).
\newblock \doi{10.1093/mnras/sty2932}

\bibitem{Foucart2016}
{Foucart}, F., {Haas}, R., {Duez}, M.D., {O'Connor}, E., {Ott}, C.D.,
  {Roberts}, L., {Kidder}, L.E., {Lippuner}, J., {Pfeiffer}, H.P., {Scheel},
  M.A.: {Low mass binary neutron star mergers: Gravitational waves and neutrino
  emission}.
\newblock Phys. Rev. D \textbf{93}(4), 044019 (2016).
\newblock \doi{10.1103/PhysRevD.93.044019}

\bibitem{Fox2005}
Fox, D.B., Frail, D.A., Price, P.A., Kulkarni, S.R., Berger, E., Piran, T.,
  Soderberg, A.M., Cenko, S.B., Cameron, P.B., Gal-Yam, A., et~al.: The afterglow of {GRB050709} and the nature of
  the short-hard gamma-ray bursts.
\newblock Nature \textbf{437}, 845 (2005)

\bibitem{Fujibayashi2018}
{Fujibayashi}, S., {Kiuchi}, K., {Nishimura}, N., {Sekiguchi}, Y., {Shibata},
  M.: {Mass Ejection from the Remnant of a Binary Neutron Star Merger:
  Viscous-radiation Hydrodynamics Study}.
\newblock Astrophys. J. \textbf{860}(1), 64 (2018).
\newblock \doi{10.3847/1538-4357/aabafd}

\bibitem{Fujibayashi2020}
{Fujibayashi}, S., {Shibata}, M., {Wanajo}, S., {Kiuchi}, K., {Kyutoku}, K.,
  {Sekiguchi}, Y.: {Mass ejection from disks surrounding a low-mass black hole:
  Viscous neutrino-radiation hydrodynamics simulation in full general
  relativity}.
\newblock Phys. Rev. D \textbf{101}(8), 083029 (2020).
\newblock \doi{10.1103/PhysRevD.101.083029}

\bibitem{Galama1998}
{Galama}, T.J., {Vreeswijk}, P.M., {van Paradijs}, J., {Kouveliotou}, C.,
  {Augusteijn}, T., {B{\"o}hnhardt}, H., {Brewer}, J.P., {Doublier}, V.,
  {Gonzalez}, J.F., {Leibundgut}, B., et~al.: {An unusual supernova in the error box of the {$\gamma$}-ray
  burst of 25 April 1998}.
\newblock Nature \textbf{395}, 670 (1998).
\newblock \doi{10.1038/27150}

\bibitem{Gao2006}
{Gao}, W.H., {Fan}, Y.Z.: {Short-living Supermassive Magnetar Model for the
  Early X-ray Flares Following Short GRBs}.
\newblock Chinese J. Astron. Astrophys. \textbf{6}, 513 (2006).
\newblock \doi{10.1088/1009-9271/6/5/01}

\bibitem{Gehrels2004-Swift}
{Gehrels}, N., {Chincarini}, G., {Giommi}, P., {Mason}, K.O., {Nousek}, J.A.,
  {Wells}, A.A., {White}, N.E., {Barthelmy}, S.D., {Burrows}, D.N., {Cominsky},
  L.R., et~al.: {The Swift Gamma-Ray
  Burst Mission}.
\newblock Astrophys. J. \textbf{611}, 1005 (2004).
\newblock \doi{10.1086/422091}

\bibitem{Gehrels2005}
{Gehrels}, N., {Sarazin}, C.L., {O'Brien}, P.T., {Zhang}, B., {Barbier}, L.,
  {Barthelmy}, S.D., {Blustin}, A., {Burrows}, D.N., {Cannizzo}, J.,
  {Cummings}, J.R., et~al.: {A short {$\gamma$}-ray burst apparently associated with an
  elliptical galaxy at redshift z = 0.225}.
\newblock Nature \textbf{437}, 851 (2005).
\newblock \doi{10.1038/nature04142}

\bibitem{Ghirlanda2019}
{Ghirlanda}, G., {Salafia}, O.S., {Paragi}, Z., {Giroletti}, M., {Yang}, J.,
  {Marcote}, B., {Blanchard}, J., {Agudo}, I., {An}, T., {Bernardini}, M.G., et~al.: {Compact
  radio emission indicates a structured jet was produced by a binary neutron
  star merger}.
\newblock Science \textbf{363}(6430), 968 (2019).
\newblock \doi{10.1126/science.aau8815}

\bibitem{Giacomazzo2009}
{Giacomazzo}, B., {Rezzolla}, L., {Baiotti}, L.: {Can magnetic fields be
  detected during the inspiral of binary neutron stars?}
\newblock Mon. Not. R. Astron. Soc. \textbf{399}, L164 (2009).
\newblock \doi{10.1111/j.1745-3933.2009.00745.x}

\bibitem{Giacomazzo2011}
{Giacomazzo}, B., {Rezzolla}, L., {Baiotti}, L.: {Accurate evolutions of
  inspiralling and magnetized neutron stars: Equal-mass binaries}.
\newblock Phys. Rev. D \textbf{83}(4), 044014 (2011).
\newblock \doi{10.1103/PhysRevD.83.044014}

\bibitem{Giacomazzo2015}
{Giacomazzo}, B., {Zrake}, J., {Duffell}, P., {MacFadyen}, A.I., {Perna}, R.:
  {Producing Magnetar Magnetic Fields in the Merger of Binary Neutron Stars}.
\newblock Astrophys. J. \textbf{809}(1), 39 (2015).
\newblock \doi{10.1088/0004-637X/809/1/39}

\bibitem{Goldstein2017}
{Goldstein}, A., {Veres}, P., {Burns}, E., {Briggs}, M.S., {Hamburg}, R.,
  {Kocevski}, D., {Wilson-Hodge}, C.A., {Preece}, R.D., {Poolakkil}, S.,
  {Roberts}, O.J., et~al.: {An Ordinary Short
  Gamma-Ray Burst with Extraordinary Implications: Fermi-GBM Detection of GRB
  170817A}.
\newblock Astrophys. J. Lett. \textbf{848}, L14 (2017).
\newblock \doi{10.3847/2041-8213/aa8f41}

\bibitem{Grossman2014}
{Grossman}, D., {Korobkin}, O., {Rosswog}, S., {Piran}, T.: {The long-term
  evolution of neutron star merger remnants - II. Radioactively powered
  transients}.
\newblock Mon. Not. R. Astron. Soc. \textbf{439}(1), 757 (2014).
\newblock \doi{10.1093/mnras/stt2503}

\bibitem{Hallinan2017}
{Hallinan}, G., {Corsi}, A., {Mooley}, K.P., {Hotokezaka}, K., {Nakar}, E.,
  {Kasliwal}, M.M., {Kaplan}, D.L., {Frail}, D.A., {Myers}, S.T., {Murphy}, T., et~al.: 
  {A radio counterpart to a neutron star merger}.
\newblock Science \textbf{358}, 1579 (2017).
\newblock \doi{10.1126/science.aap9855}

\bibitem{Hanauske2017}
{Hanauske}, M., {Takami}, K., {Bovard}, L., {Rezzolla}, L., {Font}, J.A.,
  {Galeazzi}, F., {St{\"o}cker}, H.: {Rotational properties of hypermassive
  neutron stars from binary mergers}.
\newblock Phys. Rev. D \textbf{96}(4), 043004 (2017).
\newblock \doi{10.1103/PhysRevD.96.043004}

\bibitem{Hinderer2019}
{Hinderer}, T., {Nissanke}, S., {Foucart}, F., {Hotokezaka}, K., {Vincent}, T.,
  {Kasliwal}, M., {Schmidt}, P., {Williamson}, A.R., {Nichols}, D.A., {Duez},
  M.D., et~al.: {Distinguishing the
  nature of comparable-mass neutron star binary systems with multimessenger
  observations: GW170817 case study}.
\newblock Phys. Rev. D \textbf{100}(6), 063021 (2019).
\newblock \doi{10.1103/PhysRevD.100.063021}

\bibitem{Hotokezaka2018}
{Hotokezaka}, K., {Beniamini}, P., {Piran}, T.: {Neutron star mergers as sites
  of r-process nucleosynthesis and short gamma-ray bursts}.
\newblock Int. J. of Mod. Phys. D \textbf{27}(13), 1842005 (2018).
\newblock \doi{10.1142/S0218271818420051}

\bibitem{Hotokezaka2013b}
{Hotokezaka}, K., {Kiuchi}, K., {Kyutoku}, K., {Muranushi}, T., {Sekiguchi},
  Y.i., {Shibata}, M., {Taniguchi}, K.: {Remnant massive neutron stars of
  binary neutron star mergers: Evolution process and gravitational waveform}.
\newblock Phys. Rev. D \textbf{88}(4), 044026 (2013).
\newblock \doi{10.1103/PhysRevD.88.044026}

\bibitem{Hotokezaka2013}
{Hotokezaka}, K., {Kiuchi}, K., {Kyutoku}, K., {Okawa}, H., {Sekiguchi}, Y.i.,
  {Shibata}, M., {Taniguchi}, K.: {Mass ejection from the merger of binary
  neutron stars}.
\newblock Phys. Rev. D \textbf{87}(2), 024001 (2013).
\newblock \doi{10.1103/PhysRevD.87.024001}

\bibitem{Just2015}
{Just}, O., {Bauswein}, A., {Ardevol Pulpillo}, R., {Goriely}, S., {Janka},
  H.T.: {Comprehensive nucleosynthesis analysis for ejecta of compact binary
  mergers}.
\newblock Mon. Not. R. Astron. Soc. \textbf{448}(1), 541 (2015).
\newblock \doi{10.1093/mnras/stv009}

\bibitem{Just2016}
{Just}, O., {Obergaulinger}, M., {Janka}, H.T., {Bauswein}, A., {Schwarz}, N.:
  {Neutron-star Merger Ejecta as Obstacles to Neutrino-powered Jets of
  Gamma-Ray Bursts}.
\newblock Astrophys. J. Lett. \textbf{816}, L30 (2016).
\newblock \doi{10.3847/2041-8205/816/2/L30}

\bibitem{Kasen2013}
{Kasen}, D., {Badnell}, N.R., {Barnes}, J.: {Opacities and Spectra of the
  r-process Ejecta from Neutron Star Mergers}.
\newblock Astrophys. J. \textbf{774}(1), 25 (2013).
\newblock \doi{10.1088/0004-637X/774/1/25}

\bibitem{Kasen2017}
{Kasen}, D., {Metzger}, B., {Barnes}, J., {Quataert}, E., {Ramirez-Ruiz}, E.:
  {Origin of the heavy elements in binary neutron-star mergers from a
  gravitational-wave event}.
\newblock Nature \textbf{551}(7678), 80 (2017).
\newblock \doi{10.1038/nature24453}

\bibitem{Kastaun2016}
{Kastaun}, W., {Ciolfi}, R., {Giacomazzo}, B.: {Structure of stable binary
  neutron star merger remnants: A case study}.
\newblock Phys. Rev. D \textbf{94}(4), 044060 (2016).
\newblock \doi{10.1103/PhysRevD.94.044060}

\bibitem{Kastaun2015}
Kastaun, W., Galeazzi, F.: Properties of hypermassive neutron stars formed in
  mergers of spinning binaries.
\newblock Phys. Rev. D \textbf{91}, 064027 (2015).
\newblock \doi{10.1103/PhysRevD.91.064027}

\bibitem{Kawaguchi2018}
{Kawaguchi}, K., {Shibata}, M., {Tanaka}, M.: {Radiative Transfer Simulation
  for the Optical and Near-infrared Electromagnetic Counterparts to GW170817}.
\newblock Astrophys. J. Lett. \textbf{865}(2), L21 (2018).
\newblock \doi{10.3847/2041-8213/aade02}

\bibitem{Kawamura2016}
{Kawamura}, T., {Giacomazzo}, B., {Kastaun}, W., {Ciolfi}, R., {Endrizzi}, A.,
  {Baiotti}, L., {Perna}, R.: Binary neutron star mergers and short gamma-ray
  bursts: Effects of magnetic field orientation, equation of state, and mass
  ratio.
\newblock Phys. Rev. D \textbf{94}, 064012 (2016).
\newblock \doi{10.1103/PhysRevD.94.064012}

\bibitem{Kiuchi2015}
{Kiuchi}, K., {Cerd{\'a}-Dur{\'a}n}, P., {Kyutoku}, K., {Sekiguchi}, Y.,
  {Shibata}, M.: {Efficient magnetic-field amplification due to the
  Kelvin-Helmholtz instability in binary neutron star mergers}.
\newblock Phys. Rev. D \textbf{92}(12), 124034 (2015).
\newblock \doi{10.1103/PhysRevD.92.124034}

\bibitem{Kiuchi2018}
{Kiuchi}, K., {Kyutoku}, K., {Sekiguchi}, Y., {Shibata}, M.: {Global
  simulations of strongly magnetized remnant massive neutron stars formed in
  binary neutron star mergers}.
\newblock Phys. Rev. D \textbf{97}(12), 124039 (2018).
\newblock \doi{10.1103/PhysRevD.97.124039}

\bibitem{Kiuchi2014}
{Kiuchi}, K., {Kyutoku}, K., {Sekiguchi}, Y., {Shibata}, M., {Wada}, T.: {High
  resolution numerical relativity simulations for the merger of binary
  magnetized neutron stars}.
\newblock Phys. Rev. D \textbf{90}(4), 041502 (2014).
\newblock \doi{10.1103/PhysRevD.90.041502}

\bibitem{Kiuchi2012}
{Kiuchi}, K., {Kyutoku}, K., {Shibata}, M.: {Three-dimensional evolution of
  differentially rotating magnetized neutron stars}.
\newblock Phys. Rev. D \textbf{86}(6), 064008 (2012).
\newblock \doi{10.1103/PhysRevD.86.064008}

\bibitem{Kulkarni2005}
{Kulkarni}, S.R.: {Modeling Supernova-like Explosions Associated with Gamma-ray
  Bursts with Short Durations}.
\newblock arXiv e-prints astro-ph/0510256 (2005)

\bibitem{Kumar2015}
{Kumar}, P., {Zhang}, B.: {The physics of gamma-ray bursts and relativistic
  jets}.
\newblock Phys. Rep. \textbf{561}, 1 (2015).
\newblock \doi{10.1016/j.physrep.2014.09.008}

\bibitem{Lattimer1974}
{Lattimer}, J.M., {Schramm}, D.N.: {Black-Hole-Neutron-Star Collisions}.
\newblock Astrophys. J. Lett. \textbf{192}, L145 (1974).
\newblock \doi{10.1086/181612}

\bibitem{Lazzati2018}
{Lazzati}, D., {Perna}, R., {Morsony}, B.J., {Lopez-Camara}, D., {Cantiello},
  M., {Ciolfi}, R., {Giacomazzo}, B., {Workman}, J.C.: {Late Time Afterglow
  Observations Reveal a Collimated Relativistic Jet in the Ejecta of the Binary
  Neutron Star Merger GW170817}.
\newblock Phys. Rev. Lett. \textbf{120}(24), 241103 (2018).
\newblock \doi{10.1103/PhysRevLett.120.241103}

\bibitem{Li1998}
{Li}, L.X., {Paczy{\'n}ski}, B.: {Transient Events from Neutron Star Mergers}.
\newblock Astrophys. J. \textbf{507}, L59 (1998).
\newblock \doi{10.1086/311680}

\bibitem{Lippuner2015}
{Lippuner}, J., {Roberts}, L.F.: {r-process Lanthanide Production and Heating
  Rates in Kilonovae}.
\newblock Astrophys. J. \textbf{815}(2), 82 (2015).
\newblock \doi{10.1088/0004-637X/815/2/82}

\bibitem{Liu2008}
{Liu}, Y.T., {Shapiro}, S.L., {Etienne}, Z.B., {Taniguchi}, K.: {General
  relativistic simulations of magnetized binary neutron star mergers}.
\newblock Phys. Rev. D \textbf{78}(2), 024012 (2008).
\newblock \doi{10.1103/PhysRevD.78.024012}

\bibitem{Lorimer2008}
{Lorimer}, D.R.: {Binary and Millisecond Pulsars}.
\newblock Liv. Rev. Rel. \textbf{11}, 8 (2008).
\newblock \doi{10.12942/lrr-2008-8}

\bibitem{Lu2015}
{L{\"u}}, H.J., {Zhang}, B., {Lei}, W.H., {Li}, Y., {Lasky}, P.D.: {The
  Millisecond Magnetar Central Engine in Short GRBs}.
\newblock Astrophys. J. \textbf{805}, 89 (2015).
\newblock \doi{10.1088/0004-637X/805/2/89}

\bibitem{Lyman2018}
{Lyman}, J.D., {Lamb}, G.P., {Levan}, A.J., {Mandel}, I., {Tanvir}, N.R.,
  {Kobayashi}, S., {Gompertz}, B., {Hjorth}, J., {Fruchter}, A.S., {Kangas},
  T., et~al.: {The optical afterglow of
  the short gamma-ray burst associated with GW170817}.
\newblock Nature Astr. \textbf{2}, 751 (2018).
\newblock \doi{10.1038/s41550-018-0511-3}

\bibitem{Margutti2017}
{Margutti}, R., {Berger}, E., {Fong}, W., {Guidorzi}, C., {Alexander}, K.D.,
  {Metzger}, B.D., {Blanchard}, P.K., {Cowperthwaite}, P.S., {Chornock}, R.,
  {Eftekhari}, T., et~al.: {The Electromagnetic Counterpart of the
  Binary Neutron Star Merger LIGO/Virgo GW170817. V. Rising X-Ray Emission from
  an Off-axis Jet}.
\newblock Astrophys. J. Lett. \textbf{848}, L20 (2017).
\newblock \doi{10.3847/2041-8213/aa9057}

\bibitem{Martin2015}
{Martin}, D., {Perego}, A., {Arcones}, A., {Thielemann}, F.K., {Korobkin}, O.,
  {Rosswog}, S.: {Neutrino-driven Winds in the Aftermath of a Neutron Star
  Merger: Nucleosynthesis and Electromagnetic Transients}.
\newblock Astrophys. J. \textbf{813}(1), 2 (2015).
\newblock \doi{10.1088/0004-637X/813/1/2}

\bibitem{Metzger2019LRR}
{Metzger}, B.D.: {Kilonovae}.
\newblock Liv. Rev. Rel. \textbf{23}(1), 1 (2019).
\newblock \doi{10.1007/s41114-019-0024-0}

\bibitem{Metzger2010}
{Metzger}, B.D., {Mart{\'{\i}}nez-Pinedo}, G., {Darbha}, S., {Quataert}, E.,
  {Arcones}, A., {Kasen}, D., {Thomas}, R., {Nugent}, P., {Panov}, I.V.,
  {Zinner}, N.T.: {Electromagnetic counterparts of compact object mergers
  powered by the radioactive decay of r-process nuclei}.
\newblock Mon. Not. R. Astron. Soc. \textbf{406}, 2650 (2010).
\newblock \doi{10.1111/j.1365-2966.2010.16864.x}

\bibitem{Metzger2014}
{Metzger}, B.D., {Piro}, A.L.: {Optical and X-ray emission from stable
  millisecond magnetars formed from the merger of binary neutron stars}.
\newblock Mon. Not. R. Astron. Soc. \textbf{439}, 3916 (2014).
\newblock \doi{10.1093/mnras/stu247}

\bibitem{Metzger2008}
{Metzger}, B.D., {Quataert}, E., {Thompson}, T.A.: {Short-duration gamma-ray
  bursts with extended emission from protomagnetar spin-down}.
\newblock Mon. Not. R. Astron. Soc. \textbf{385}, 1455 (2008).
\newblock \doi{10.1111/j.1365-2966.2008.12923.x}

\bibitem{Metzger2018}
{Metzger}, B.D., {Thompson}, T.A., {Quataert}, E.: {A Magnetar Origin for the
  Kilonova Ejecta in GW170817}.
\newblock Astrophys. J. Lett. \textbf{856}(2), 101 (2018).
\newblock \doi{10.3847/1538-4357/aab095}

\bibitem{Meyer1989}
{Meyer}, B.S.: {Decompression of Initially Cold Neutron Star Matter: A
  Mechanism for the r-Process?}
\newblock Astrophys. J. \textbf{343}, 254 (1989).
\newblock \doi{10.1086/167702}

\bibitem{Mochkovitch1993}
{Mochkovitch}, R., {Hernanz}, M., {Isern}, J., {Martin}, X.: {Gamma-ray bursts
  as collimated jets from neutron star/black hole mergers}.
\newblock Nature \textbf{361}, 236 (1993).
\newblock \doi{10.1038/361236a0}

\bibitem{Mooley2018b}
{Mooley}, K.P., {Deller}, A.T., {Gottlieb}, O., {Nakar}, E., {Hallinan}, G.,
  {Bourke}, S., {Frail}, D.A., {Horesh}, A., {Corsi}, A., {Hotokezaka}, K.:
  {Superluminal motion of a relativistic jet in the neutron-star merger
  GW170817}.
\newblock Nature \textbf{561}, 355 (2018).
\newblock \doi{10.1038/s41586-018-0486-3}

\bibitem{Mooley2018a}
{Mooley}, K.P., {Nakar}, E., {Hotokezaka}, K., {Hallinan}, G., {Corsi}, A.,
  {Frail}, D.A., {Horesh}, A., {Murphy}, T., {Lenc}, E., {Kaplan}, D.L., et~al.: {A
  mildly relativistic wide-angle outflow in the neutron-star merger event
  GW170817}.
\newblock Nature \textbf{554}, 207 (2018).
\newblock \doi{10.1038/nature25452}

\bibitem{Most2019}
{Most}, E.R., {Papenfort}, L.J., {Rezzolla}, L.: {Beyond second-order
  convergence in simulations of magnetized binary neutron stars with realistic
  microphysics}.
\newblock Mon. Not. R. Astron. Soc. \textbf{490}(3), 3588 (2019).
\newblock \doi{10.1093/mnras/stz2809}

\bibitem{Moesta2020}
{M{\"o}sta}, P., {Radice}, D., {Haas}, R., {Schnetter}, E., {Bernuzzi}, S.: {A
  magnetar engine for short GRBs and kilonovae}.
\newblock arXiv e-prints arXiv:2003.06043 (2020)

\bibitem{Narayan1992}
{Narayan}, R., {Paczynski}, B., {Piran}, T.: {Gamma-ray bursts as the death
  throes of massive binary stars}.
\newblock Astrophys. J. Lett. \textbf{395}, L83 (1992).
\newblock \doi{10.1086/186493}

\bibitem{Nathanail2018}
{Nathanail}, A.: {Binary Neutron Star and Short Gamma-Ray Burst Simulations in
  Light of GW170817}.
\newblock Galaxies \textbf{6}(4), 119 (2018).
\newblock \doi{10.3390/galaxies6040119}

\bibitem{Nedora2019}
{Nedora}, V., {Bernuzzi}, S., {Radice}, D., {Perego}, A., {Endrizzi}, A.,
  {Ortiz}, N.: {Spiral-wave Wind for the Blue Kilonova}.
\newblock Astrophys. J. Lett. \textbf{886}(2), L30 (2019).
\newblock \doi{10.3847/2041-8213/ab5794}

\bibitem{Neilsen2014}
{Neilsen}, D., {Liebling}, S.L., {Anderson}, M., {Lehner}, L., {O'Connor}, E.,
  {Palenzuela}, C.: {Magnetized neutron stars with realistic equations of state
  and neutrino cooling}.
\newblock Phys. Rev. D \textbf{89}(10), 104029 (2014).
\newblock \doi{10.1103/PhysRevD.89.104029}

\bibitem{Oganesyan2020}
{Oganesyan}, G., {Ascenzi}, S., {Branchesi}, M., {Salafia}, O.S., {Dall'Osso},
  S., {Ghirlanda}, G.: {Structured Jets and X-Ray Plateaus in Gamma-Ray Burst
  Phenomena}.
\newblock Astrophys. J. \textbf{893}(2), 88 (2020).
\newblock \doi{10.3847/1538-4357/ab8221}

\bibitem{Paczynski1986}
{Paczynski}, B.: {Gamma-ray bursters at cosmological distances}.
\newblock Astrophys. J. Lett. \textbf{308}, L43 (1986).
\newblock \doi{10.1086/184740}

\bibitem{Palenzuela2013}
{Palenzuela}, C., {Lehner}, L., {Ponce}, M., {Liebling}, S.L., {Anderson}, M.,
  {Neilsen}, D., {Motl}, P.: {Electromagnetic and Gravitational Outputs from
  Binary-Neutron-Star Coalescence}.
\newblock Phys. Rev. Lett. \textbf{111}(6), 061105 (2013).
\newblock \doi{10.1103/PhysRevLett.111.061105}

\bibitem{Palenzuela2015}
{Palenzuela}, C., {Liebling}, S.L., {Neilsen}, D., {Lehner}, L., {Caballero},
  O.L., {O'Connor}, E., {Anderson}, M.: {Effects of the microphysical equation
  of state in the mergers of magnetized neutron stars with neutrino cooling}.
\newblock Phys. Rev. D \textbf{92}(4), 044045 (2015).
\newblock \doi{10.1103/PhysRevD.92.044045}

\bibitem{Paschalidis2015}
{Paschalidis}, V., {Ruiz}, M., {Shapiro}, S.L.: {Relativistic Simulations of
  Black Hole-Neutron Star Coalescence: The Jet Emerges}.
\newblock Astrophys. J. Lett. \textbf{806}, L14 (2015).
\newblock \doi{10.1088/2041-8205/806/1/L14}

\bibitem{Paschalidis2017}
{Paschalidis}, V.: {General relativistic simulations of compact binary mergers
  as engines for short gamma-ray bursts}.
\newblock Class. Quantum Grav. \textbf{34}(8), 084002 (2017).
\newblock \doi{10.1088/1361-6382/aa61ce}

\bibitem{Paschalidis2017LRR}
{Paschalidis}, V., {Stergioulas}, N.: {Rotating stars in relativity}.
\newblock Liv. Rev. Rel. \textbf{20}(1), 7 (2017).
\newblock \doi{10.1007/s41114-017-0008-x}

\bibitem{Paschalidis2019}
{Paschalidis}, V., {Ruiz}, M.: {Are fast radio bursts the most likely
  electromagnetic counterpart of neutron star mergers resulting in prompt
  collapse?}
\newblock Phys. Rev. D \textbf{100}(4), 043001 (2019).
\newblock \doi{10.1103/PhysRevD.100.043001}

\bibitem{Perego2017b}
{Perego}, A., {Radice}, D., {Bernuzzi}, S.: {AT 2017gfo: An Anisotropic and
  Three-component Kilonova Counterpart of GW170817}.
\newblock Astrophys. J. Lett. \textbf{850}(2), L37 (2017).
\newblock \doi{10.3847/2041-8213/aa9ab9}

\bibitem{Perego2014}
{Perego}, A., {Rosswog}, S., {Cabez{\'o}n}, R.M., {Korobkin}, O.,
  {K{\"a}ppeli}, R., {Arcones}, A., {Liebend{\"o}rfer}, M.: {Neutrino-driven
  winds from neutron star merger remnants}.
\newblock Mon. Not. R. Astron. Soc. \textbf{443}(4), 3134 (2014).
\newblock \doi{10.1093/mnras/stu1352}

\bibitem{Perego2017a}
{Perego}, A., {Yasin}, H., {Arcones}, A.: {Neutrino pair annihilation above
  merger remnants: implications of a long-lived massive neutron star}.
\newblock J. Phys. G Nucl. Phys. \textbf{44}(8), 084007 (2017).
\newblock \doi{10.1088/1361-6471/aa7bdc}

\bibitem{Pian2017}
{Pian}, E., {D'Avanzo}, P., {Benetti}, S., {Branchesi}, M., {Brocato}, E.,
  {Campana}, S., {Cappellaro}, E., {Covino}, S., {D'Elia}, V., {Fynbo}, J.P.U., 
  et~al.: {Spectroscopic identification of r-process
  nucleosynthesis in a double neutron-star merger}.
\newblock Nature \textbf{551}(7678), 67 (2017).
\newblock \doi{10.1038/nature24298}

\bibitem{piran2004}
{Piran}, T.: {The physics of gamma-ray bursts}.
\newblock Rev. Mod. Phys. \textbf{76}, 1143 (2004).
\newblock \doi{10.1103/RevModPhys.76.1143}

\bibitem{Ponce2014}
{Ponce}, M., {Palenzuela}, C., {Lehner}, L., {Liebling}, S.L.: {Interaction of
  misaligned magnetospheres in the coalescence of binary neutron stars}.
\newblock Phys. Rev. D \textbf{90}(4), 044007 (2014).
\newblock \doi{10.1103/PhysRevD.90.044007}

\bibitem{Price2006}
{Price}, D.J., {Rosswog}, S.: {Producing Ultrastrong Magnetic Fields in Neutron
  Star Mergers}.
\newblock Science \textbf{312}(5774), 719 (2006).
\newblock \doi{10.1126/science.1125201}

\bibitem{Radice2017}
{Radice}, D.: {General-relativistic Large-eddy Simulations of Binary Neutron
  Star Mergers}.
\newblock Astrophys. J. Lett. \textbf{838}(1), L2 (2017).
\newblock \doi{10.3847/2041-8213/aa6483}

\bibitem{Radice2016}
{Radice}, D., {Galeazzi}, F., {Lippuner}, J., {Roberts}, L.F., {Ott}, C.D.,
  {Rezzolla}, L.: {Dynamical mass ejection from binary neutron star mergers}.
\newblock Mon. Not. R. Astron. Soc. \textbf{460}(3), 3255 (2016).
\newblock \doi{10.1093/mnras/stw1227}

\bibitem{Raithel2019}
{Raithel}, C.A.: {Constraints on the neutron star equation of state from
  GW170817}.
\newblock European Phys. J. A \textbf{55}(5), 80 (2019).
\newblock \doi{10.1140/epja/i2019-12759-5}

\bibitem{Rea2010}
{Rea}, N., {Esposito}, P., {Turolla}, R., {Israel}, G.L., {Zane}, S., {Stella},
  L., {Mereghetti}, S., {Tiengo}, A., {G{\"o}tz}, D., {G{\"o}{\u{g}}{\"u}s},
  E., et~al.: {A Low-Magnetic-Field Soft Gamma Repeater}.
\newblock Science \textbf{330}(6006), 944 (2010).
\newblock \doi{10.1126/science.1196088}

\bibitem{Rezzolla2011}
{Rezzolla}, L., {Giacomazzo}, B., {Baiotti}, L., {Granot}, J., {Kouveliotou},
  C., {Aloy}, M.A.: {The missing link: Merging neutron stars naturally produce
  jet-like structures and can power short Gamma-Ray Bursts}.
\newblock Astrophys. J. Lett. \textbf{732}(11), L6 (2011).
\newblock \doi{10.1088/2041-8205/732/1/L6}

\bibitem{Rosswog2005}
{Rosswog}, S.: {Mergers of Neutron Star-Black Hole Binaries with Small Mass
  Ratios: Nucleosynthesis, Gamma-Ray Bursts, and Electromagnetic Transients}.
\newblock Astrophys. J. \textbf{634}, 1202 (2005).
\newblock \doi{10.1086/497062}

\bibitem{Rowlinson2013}
{Rowlinson}, A., {O'Brien}, P.T., {Metzger}, B.D., {Tanvir}, N.R., {Levan},
  A.J.: {Signatures of magnetar central engines in short GRB light curves}.
\newblock Mon. Not. R. Astron. Soc. \textbf{430}, 1061 (2013).
\newblock \doi{10.1093/mnras/sts683}

\bibitem{Rowlinson2010}
{Rowlinson}, A., {O'Brien}, P.T., {Tanvir}, N.R., {Zhang}, B., {Evans}, P.A.,
  {Lyons}, N., {Levan}, A.J., {Willingale}, R., {Page}, K.L., {Onal}, O., et~al.: {The
  unusual X-ray emission of the short Swift GRB 090515: evidence for the
  formation of a magnetar?}
\newblock Mon. Not. R. Astron. Soc. \textbf{409}, 531 (2010).
\newblock \doi{10.1111/j.1365-2966.2010.17354.x}

\bibitem{Ruffert1999}
Ruffert, M., Janka, H.T.: Gamma-ray bursts from accreting black holes in
  neutron star mergers.
\newblock Astron. Astrophys. \textbf{344}, 573 (1999)

\bibitem{Ruiz2016}
{Ruiz}, M., {Lang}, R.N., {Paschalidis}, V., {Shapiro}, S.L.: {Binary Neutron
  Star Mergers: A Jet Engine for Short Gamma-Ray Bursts}.
\newblock Astrophys. J. Lett. \textbf{824}, L6 (2016).
\newblock \doi{10.3847/2041-8205/824/1/L6}

\bibitem{Ruiz2017}
{Ruiz}, M., {Shapiro}, S.L.: {General relativistic magnetohydrodynamics
  simulations of prompt-collapse neutron star mergers: The absence of jets}.
\newblock Phys. Rev. D \textbf{96}(8), 084063 (2017).
\newblock \doi{10.1103/PhysRevD.96.084063}

\bibitem{Ruiz2018}
{Ruiz}, M., {Shapiro}, S.L., {Tsokaros}, A.: {GW170817, general relativistic
  magnetohydrodynamic simulations, and the neutron star maximum mass}.
\newblock Phys. Rev. D \textbf{97}, 021501 (2018).
\newblock \doi{10.1103/PhysRevD.97.021501}

\bibitem{Ruiz2019}
{Ruiz}, M., {Tsokaros}, A., {Paschalidis}, V., {Shapiro}, S.L.: {Effects of
  spin on magnetized binary neutron star mergers and jet launching}.
\newblock Phys. Rev. D \textbf{99}(8), 084032 (2019).
\newblock \doi{10.1103/PhysRevD.99.084032}

\bibitem{Ruiz2020}
{Ruiz}, M., {Tsokaros}, A., {Shapiro}, S.L.: {Magnetohydrodynamic simulations
  of binary neutron star mergers in general relativity: Effects of magnetic
  field orientation on jet launching}.
\newblock Phys. Rev. D \textbf{101}(6), 064042 (2020).
\newblock \doi{10.1103/PhysRevD.101.064042}

\bibitem{Salafia2019}
{Salafia}, O.S., {Ghirlanda}, G., {Ascenzi}, S., {Ghisellini}, G.: {On-axis
  view of GRB 170817A}.
\newblock Astron. Astrophys. \textbf{628}, A18 (2019).
\newblock \doi{10.1051/0004-6361/201935831}

\bibitem{Savchenko2017}
{Savchenko}, V., {Ferrigno}, C., {Kuulkers}, E., {Bazzano}, A., {Bozzo}, E.,
  {Brandt}, S., {Chenevez}, J., {Courvoisier}, T.J.L., {Diehl}, R., {Domingo},
  A., et~al.: {INTEGRAL
  Detection of the First Prompt Gamma-Ray Signal Coincident with the
  Gravitational-wave Event GW170817}.
\newblock Astrophys. J. Lett. \textbf{848}, L15 (2017).
\newblock \doi{10.3847/2041-8213/aa8f94}

\bibitem{Sekiguchi2011}
{Sekiguchi}, Y., {Kiuchi}, K., {Kyutoku}, K., {Shibata}, M.: {Gravitational
  Waves and Neutrino Emission from the Merger of Binary Neutron Stars}.
\newblock Phys. Rev. Lett. \textbf{107}(5), 051102 (2011).
\newblock \doi{10.1103/PhysRevLett.107.051102}

\bibitem{Sekiguchi2016}
{Sekiguchi}, Y., {Kiuchi}, K., {Kyutoku}, K., {Shibata}, M., {Taniguchi}, K.:
  {Dynamical mass ejection from the merger of asymmetric binary neutron stars:
  Radiation-hydrodynamics study in general relativity}.
\newblock Phys. Rev. D \textbf{93}(12), 124046 (2016).
\newblock \doi{10.1103/PhysRevD.93.124046}

\bibitem{Shibata2019}
{Shibata}, M., {Hotokezaka}, K.: {Merger and Mass Ejection of Neutron Star
  Binaries}.
\newblock Ann. Rev. Nuc. Part. Sci. \textbf{69}, 41 (2019).
\newblock \doi{10.1146/annurev-nucl-101918-023625}

\bibitem{Shibata2017b}
{Shibata}, M., {Kiuchi}, K.: {Gravitational waves from remnant massive neutron
  stars of binary neutron star merger: Viscous hydrodynamics effects}.
\newblock Phys. Rev. D \textbf{95}(12), 123003 (2017).
\newblock \doi{10.1103/PhysRevD.95.123003}

\bibitem{Shibata2017a}
{Shibata}, M., {Kiuchi}, K., {Sekiguchi}, Y.i.: {General relativistic viscous
  hydrodynamics of differentially rotating neutron stars}.
\newblock Phys. Rev. D \textbf{95}(8), 083005 (2017).
\newblock \doi{10.1103/PhysRevD.95.083005}

\bibitem{Shibata2011}
{Shibata}, M., {Suwa}, Y., {Kiuchi}, K., {Ioka}, K.: {Afterglow of a Binary
  Neutron Star Merger}.
\newblock Astrophys. J. Lett. \textbf{734}(2), L36 (2011).
\newblock \doi{10.1088/2041-8205/734/2/L36}

\bibitem{Shibata2005}
{Shibata}, M., {Taniguchi}, K., {Ury{\={u}}}, K.: {Merger of binary neutron
  stars with realistic equations of state in full general relativity}.
\newblock Phys. Rev. D \textbf{71}(8), 084021 (2005).
\newblock \doi{10.1103/PhysRevD.71.084021}

\bibitem{Shibata2000}
{Shibata}, M., {Ury{\={u}}}, K.{\={o}}.: {Simulation of merging binary neutron
  stars in full general relativity: {\ensuremath{\Gamma}}=2 case}.
\newblock Phys. Rev. D \textbf{61}(6), 064001 (2000).
\newblock \doi{10.1103/PhysRevD.61.064001}

\bibitem{Siegel2019}
{Siegel}, D.M.: {GW170817 -the first observed neutron star merger and its
  kilonova: Implications for the astrophysical site of the r-process}.
\newblock European Phys. J. A \textbf{55}(11), 203 (2019).
\newblock \doi{10.1140/epja/i2019-12888-9}

\bibitem{Siegel2016a}
{Siegel}, D.M., {Ciolfi}, R.: {Electromagnetic Emission from Long-lived Binary
  Neutron Star Merger Remnants. I. Formulation of the Problem}.
\newblock Astrophys. J. \textbf{819}, 14 (2016).
\newblock \doi{10.3847/0004-637X/819/1/14}

\bibitem{Siegel2016b}
{Siegel}, D.M., {Ciolfi}, R.: {Electromagnetic Emission from Long-lived Binary
  Neutron Star Merger Remnants. II. Lightcurves and Spectra}.
\newblock Astrophys. J. \textbf{819}, 15 (2016).
\newblock \doi{10.3847/0004-637X/819/1/15}

\bibitem{Siegel2013}
{Siegel}, D.M., {Ciolfi}, R., {Harte}, A.I., {Rezzolla}, L.: {Magnetorotational
  instability in relativistic hypermassive neutron stars}.
\newblock Phys. Rev. D(R) \textbf{87}(12), 121302 (2013).
\newblock \doi{10.1103/PhysRevD.87.121302}

\bibitem{Siegel2014}
{Siegel}, D.M., {Ciolfi}, R., {Rezzolla}, L.: {Magnetically Driven Winds from
  Differentially Rotating Neutron Stars and X-Ray Afterglows of Short Gamma-Ray
  Bursts}.
\newblock Astrophys. J. Lett. \textbf{785}(1), L6 (2014).
\newblock \doi{10.1088/2041-8205/785/1/L6}

\bibitem{Siegel2017}
{Siegel}, D.M., {Metzger}, B.D.: {Three-Dimensional General-Relativistic
  Magnetohydrodynamic Simulations of Remnant Accretion Disks from Neutron Star
  Mergers: Outflows and r -Process Nucleosynthesis}.
\newblock Phys. Rev. Lett. \textbf{119}(23), 231102 (2017).
\newblock \doi{10.1103/PhysRevLett.119.231102}

\bibitem{Siegel2018}
{Siegel}, D.M., {Metzger}, B.D.: {Three-dimensional GRMHD Simulations of
  Neutrino-cooled Accretion Disks from Neutron Star Mergers}.
\newblock Astrophys. J. \textbf{858}(1), 52 (2018).
\newblock \doi{10.3847/1538-4357/aabaec}

\bibitem{Smartt2017}
{Smartt}, S.J., {Chen}, T.W., {Jerkstrand}, A., {Coughlin}, M., {Kankare}, E.,
  {Sim}, S.A., {Fraser}, M., {Inserra}, C., {Maguire}, K., {Chambers}, K.C.,
  et~al.: {A kilonova as the electromagnetic
  counterpart to a gravitational-wave source}.
\newblock Nature \textbf{551}(7678), 75 (2017).
\newblock \doi{10.1038/nature24303}

\bibitem{Stergioulas2011}
{Stergioulas}, N., {Bauswein}, A., {Zagkouris}, K., {Janka}, H.T.:
  {Gravitational waves and non-axisymmetric oscillation modes in mergers of
  compact object binaries}.
\newblock Mon. Not. R. Astron. Soc. \textbf{418}(1), 427 (2011).
\newblock \doi{10.1111/j.1365-2966.2011.19493.x}

\bibitem{Symbalisty1982}
{Symbalisty}, E., {Schramm}, D.N.: {Neutron Star Collisions and the r-Process}.
\newblock Astrophys. Lett. \textbf{22}, 143 (1982)

\bibitem{Takami2014}
{Takami}, K., {Rezzolla}, L., {Baiotti}, L.: {Constraining the Equation of
  State of Neutron Stars from Binary Mergers}.
\newblock Phys. Rev. Lett. \textbf{113}(9), 091104 (2014).
\newblock \doi{10.1103/PhysRevLett.113.091104}

\bibitem{Tanaka2013}
{Tanaka}, M., {Hotokezaka}, K.: {Radiative Transfer Simulations of Neutron Star
  Merger Ejecta}.
\newblock Astrophys. J. \textbf{775}(2), 113 (2013).
\newblock \doi{10.1088/0004-637X/775/2/113}

\bibitem{Tanvir2013}
{Tanvir}, N.R., {Levan}, A.J., {Fruchter}, A.S., {Hjorth}, J., {Hounsell},
  R.A., {Wiersema}, K., {Tunnicliffe}, R.L.: {A `kilonova' associated with the
  short-duration {$\gamma$}-ray burst GRB 130603B}.
\newblock Nature \textbf{500}, 547 (2013).
\newblock \doi{10.1038/nature12505}

\bibitem{Thorne1986}
{Thorne}, K.S., {Price}, R.H., {MacDonald}, D.A.: {Black holes: The membrane
  paradigm} (1986)

\bibitem{Troja2017}
{Troja}, E., {Piro}, L., {van Eerten}, H., {Wollaeger}, R.T., {Im}, M., {Fox},
  O.D., {Butler}, N.R., {Cenko}, S.B., {Sakamoto}, T., {Fryer}, C.L., et~al.: {The X-ray counterpart to the gravitational-wave event GW170817}.
\newblock Nature \textbf{551}, 71 (2017).
\newblock \doi{10.1038/nature24290}

\bibitem{Tsokaros2019}
{Tsokaros}, A., {Ruiz}, M., {Paschalidis}, V., {Shapiro}, S.L., {Ury{\={u}}},
  K.: {Effect of spin on the inspiral of binary neutron stars}.
\newblock Phys. Rev. D \textbf{100}(2), 024061 (2019).
\newblock \doi{10.1103/PhysRevD.100.024061}

\bibitem{Valenti2017}
{Valenti}, S., {Sand}, D.J., {Yang}, S., {Cappellaro}, E., {Tartaglia}, L.,
  {Corsi}, A., {Jha}, S.W., {Reichart}, D.E., {Haislip}, J., {Kouprianov}, V.:
  {The Discovery of the Electromagnetic Counterpart of GW170817: Kilonova AT
  2017gfo/DLT17ck}.
\newblock Astrophys. J. Lett. \textbf{848}(2), L24 (2017).
\newblock \doi{10.3847/2041-8213/aa8edf}

\bibitem{Villar2017}
{Villar}, V.A., {Guillochon}, J., {Berger}, E., {Metzger}, B.D.,
  {Cowperthwaite}, P.S., {Nicholl}, M., {Alexand er}, K.D., {Blanchard}, P.K.,
  {Chornock}, R., {Eftekhari}, T., et~al.: {The Combined Ultraviolet, Optical, and Near-infrared Light Curves of
  the Kilonova Associated with the Binary Neutron Star Merger GW170817: Unified
  Data Set, Analytic Models, and Physical Implications}.
\newblock Astrophys. J. Lett. \textbf{851}(1), L21 (2017).
\newblock \doi{10.3847/2041-8213/aa9c84}

\bibitem{Yang2018}
{Yang}, H., {Paschalidis}, V., {Yagi}, K., {Lehner}, L., {Pretorius}, F.,
  {Yunes}, N.: {Gravitational wave spectroscopy of binary neutron star merger
  remnants with mode stacking}.
\newblock Phys. Rev. D \textbf{97}(2), 024049 (2018).
\newblock \doi{10.1103/PhysRevD.97.024049}

\bibitem{Yu2013}
{Yu}, Y.W., {Zhang}, B., {Gao}, H.: {Bright ''Merger-nova'' from the Remnant of
  a Neutron Star Binary Merger: A Signature of a Newly Born, Massive,
  Millisecond Magnetar}.
\newblock Astrophys. J. Lett. \textbf{776}, L40 (2013).
\newblock \doi{10.1088/2041-8205/776/2/L40}

\bibitem{Zhang2001}
{Zhang}, B., {M{\'e}sz{\'a}ros}, P.: {Gamma-Ray Burst Afterglow with Continuous
  Energy Injection: Signature of a Highly Magnetized Millisecond Pulsar}.
\newblock Astrophys. J. Lett. \textbf{552}, L35 (2001).
\newblock \doi{10.1086/320255}

\end{thebibliography}

\end{document}